\documentclass[aps,pra,reprint,showpacs,superscriptaddress,groupedaddress]{revtex4-1}

\usepackage{amsmath}
\usepackage{graphicx}
\usepackage{dcolumn}
\usepackage{bm}

\begin{document}


\title{Generalized modulational instability in multimode fibers: wideband multimode parametric amplification}

\author{M. Guasoni}
\affiliation{Laboratoire Interdisciplinaire Carnot de Bourgogne, CNRS, University of Burgundy, Dijon, France}

\begin{abstract}
In this paper intermodal modulational instability (IM-MI) is analyzed in a multimode fiber where several spatial and polarization modes propagate. The coupled nonlinear Schr\"{o}dinger equations describing the modal evolution in the fiber are linearized and reduced to an eigenvalue problem. As a result, the amplification of each mode can be described by means of the eigenvalues and eigenvectors of a matrix that stores the information about the dispersion properties of the modes and the modal power distribution of the pump. Some useful analytical formulas are also provided that estimate the modal amplification as function of the system parameters. Finally, the impact of third-order dispersion and of absorbtion losses is evaluated, which reveals some surprising phenomena into the IM-MI dynamics. These outcomes generalize previous studies on bimodal-MI, related to the interaction between 2 spatial or polarization modes, to the most general case of $N>2$ interacting modes. Moreover, they pave the way towards the realization of wideband multimode parametric amplifiers.
\end{abstract}



\maketitle

\section{Introduction}\label{rif:intro}
The past 40 years have witnessed a huge increase of the transmission capacity in single-mode fibers. Every available degree of freedom has been explored, such as multiplexing in time, in wavelength, in polarization and in phase. On the other hand, nowadays single-mode fibers are gradually reaching their capacity limit of about 100 Tbit/s which is dictated by the Shannon theorem \cite{Essiambre12,Mitra01}.\\
In this scenario, multicore and multimode fibers are rapidly emerging as the ideal solution in order to fulfil the growing demand of capacity. In these fibers nonlinear pulse manipulation can be efficiently exploited for various applications \cite{Turitsyn15}. Furthermore, they allow exploring a further degree of freedom, namely the spatial, and therefore the implementation of space division multiplexing (SDM) schemes \cite{Richardson13} . In SDM each fiber core and/or mode represents an independent information channel parallel to the others,  which permits in principle to largely overcome the capacity limit of single-mode fibers.\\
A key-component for an efficient operation of SDM schemes is represented by multimode optical amplifiers
\cite{Krummrich13}, which have been the subject of intense research over the last years. Among the most important we find multimode fiber Raman amplifiers ( RA) \cite{Ryf12,Antonelli14} and multimode erbium doped fiber amplifiers (EDFA) \cite{Bai11,Jung11}.\\
An alternative fiber-based approach is represented by parametric amplification \cite{Hansryd02}, which is based on a modulational instability (MI) process.\\
Compared to RAs and EDFAs, parametric amplification guarantees more flexibility since it makes possible to tune the position of the amplified sidebands at arbitrary wavelength by means of the input pump and of the fiber parameters. Moreover, Kerr-nonlinearity is practically instantaneous, enabling ultrafast signal processing applications.\\
Intermodal-MI (IM-MI) in multimode fibers was demonstrated decades ago \cite{Stolen75} and more recently explored as a possible means for supercontinuum generation \cite{Mussot03} as well as for the development of efficient fiber parametric amplifiers \cite{Tonello06,Rishoj12} and low-cost broadband fiber sources \cite{Wright15}.\\
In all previous works, IM-MI has been analyzed by considering separately the interaction between couples of modes and studying the bimodal-MI process related to the phase-matching of each couple. With this approach, the complex IM-MI dynamics is decomposed in a set of bimodal-MI processes, which greatly simplifies its analysis. On the other hand, the presence and the influence of the non-phase-matched modes is usually neglected.\\
The main purpose of this paper is to go beyond this simplistic view and to develop a model taking into account for the interaction between all the fiber modes, which is the solely way to correctly describe the IM-MI. Furthermore we provide, for the first time to the best of our knowledge, some useful analytical estimates for the IM-MI gain and we evaluate the impact of higher-order dispersion terms and losses, which reveals some surprising phenomena into the IM-MI dynamics. The final result is a general description of the IM-MI that paves the way to the realization of wideband multimode parametric amplifiers.\\
The paper is organized as follows.\\
In Section ~\ref{rif:general} we linearize and reduce to an eigenvalue problem the coupled nonlinear Schr\"{o}dinger equations (CNLSE) describing the propagation in a multimode fiber. We focus our attention on single-core isotropic, highly-birefringent and telecommunication fibers. We finally obtain a matrix ${\bf M}$ storing all the information about the dispersion properties of the modes and the modal power distribution of the pump. This outcome generalizes previous results on bimodal-MI \cite{Seve96}, generated by the interaction between 2 spatial or polarization modes, to the most general case of $N>2$ interacting modes.\\
In Section ~\ref{rif:simulations} we find some useful analytical formulas that quantify the growth of each mode as function of the eigenvalues of ${\bf M}$ with negative imaginary part, namely the gains, and their corresponding eigenvectors. Interesting enough, several gains could coexist for the same pump-sideband detuning, giving rise to a competition between two or more amplification processes. We thus introduce the concept of dominant gain, which is the truly leader of the MI growth, and we show that different gains could be dominant at different fiber positions and for different modes.\\
In Section ~\ref{rif:inside} we provide a deeper physical insight into the IM-MI and we derive some analytical estimates for the IM-MI gain as function of the pump-sideband detuning and the system parameters. We put in evidence that IM-MI is truly associated to a phase-matched four-wave mixing process, which is the usual point of view of past works, but we also clearly show that all the fiber modes undergo amplification, and not only the phase-matched ones. The amplification of each mode can be controlled  by means of the system parameters and over a large bandwidth that increases with the number of the modes.\\
In Sections ~\ref{rif:TOD impact},~\ref{rif:pump impact} and ~\ref{rif:losses impact} we analyze the impact of third-order dispersion as well as of the pump power distribution and of the propagation losses on the IM-MI dynamics. We find that absorption losses cause a shift of the gain curves towards low frequencies and that, differently from single-mode fibers, third-order dispersion plays a substantial role and may lead to the generation of secondary MI bands in addition to conventional ones.\\
In Section ~\ref{rif:limits} we discuss the limits of validity of our model and finally, in Section ~\ref{rif:conclusions}, we resume the main outcomes and we give the concluding remarks.

\section{General theory}\label{rif:general}
Let us consider a beam, centered at the carrier angular frequency $\omega_p$, which is injected in an optical fiber and is coupled to the N propagating modes of the fiber. In the weakly guiding approximation, which is well accurate in typical silica fibers with low core-cladding refractive index difference, the modal transverse profiles $M_n(x,y)$ ( $1\leq n\leq N$) are linearly polarized and almost independent of their polarization. The  $a-$polarized component ($a=\{x,y\}$) of the total electric field ${\bf E}$ propagating in the fiber along the $z$-axis can be written as follows:

\begin{eqnarray}
E_a = \sum_{n=1}^N P_{na}(z,t) exp(i \beta_{p,na}  z +i\omega_p t) M_n(x,y) + c.c.
\label{TotalField}
\end{eqnarray}

where $P_{na}$ is the envelope of the a-polarized n-mode, which will be called $na-$mode throughout the paper, and $\beta_{p,na}\equiv \beta_{na}(\omega_p)$ is the corresponding propagation constant at the frequency $\omega_p$.\\
Starting from the Maxwell equations, with a polarization that takes into account the nonlinear cubic response of silica, the following set of coupled equations can be derived which describes the spatio-temporal dynamics of the $x$-polarized modal envelopes \cite{Agrawal13}:

\begin{flalign}
&\frac{\partial {P_{nx}}}{\partial z} = &&\nonumber\\
&- \it{v}_{nx}^{-1}\frac{\partial  P_{nx}}{\partial t}  - i\frac{\beta_{2nx}}{2}\frac{\partial^2  P_{nx}}{\partial t^2}+ \frac{\beta_{3nx}}{6}\frac{\partial^3  P_{nx}}{\partial t^3} - \alpha_{nx}P_{nx}+&&\nonumber\\
            &i\sum_{klm} c_{klmn}\left(P_{kx}P_{lx}P_{mx}^*e_{1,klmn} + 2 P_{kx}^*P_{lx}P_{mx}e_{2,klmn}   \right)+&&\nonumber\\
            &i\sum_{klm} c_{klmn}\left(P_{ky}P_{ly}P_{mx}^*e_{3,klmn} + 2 P_{ky}^*P_{ly}P_{mx}e_{4,klmn}   \right)&&
\label{TotalEquation}
\end{flalign}

An equation similar to Eq.(\ref{TotalEquation}) is valid for the $y$-polarized modal envelopes $P_{ny}$ by exchanging the labels $x\leftrightarrow y$.\\
Here $^*$ denotes the complex conjugate and $e_{r,klmn}\equiv exp(i\Delta\beta_{r,klmn} z)$, $r=\{1,2,3,4\}$, being: $\Delta\beta_{1,klmn}\equiv \beta_{p,kx}+\beta_{p,lx}-\beta_{p,mx}-\beta_{p,nx}$; $\Delta\beta_{2,klmn}\equiv -\beta_{p,kx}+\beta_{p,lx}+\beta_{p,mx}-\beta_{p,nx}$; $\Delta\beta_{3,klmn}\equiv\beta_{p,ky}+\beta_{p,ly}-\beta_{p,mx}-\beta_{p,nx}$; $\Delta\beta_{4,klmn}\equiv -\beta_{p,ky}+\beta_{p,ly}+\beta_{p,mx}-\beta_{p,nx}$.\\
Coefficients $\it{v}_{na}\equiv \partial\omega/\partial\beta_{na}|_{\omega_p}$, $\beta_{2na}\equiv \partial^2\beta_{na}/\partial\omega^2|_{\omega_p}$, $\beta_{3na}\equiv \partial^3\beta_{na}/\partial\omega^3|_{\omega_p}$ and $\alpha_{na}$ indicate respectively the group velocity, the group velocity dispersion (GVD), the third-order dispersion (TOD) and the propagation losses related to the $na$-mode at the frequency $\omega_p$.\\
 The nonlinear coupling coefficients read as $c_{klmn}\equiv(n_2 \omega_p/c)\int\int M_kM_lM_mM_n dxdy$, where $n_2$ is the nonlinear index of the fiber, $c$ is the speed of light in vacuum and all the modal profiles are normalized so that the area $\int\int M_n^2 dxdy = 1$.   \\
Equation (\ref{TotalEquation}) is valid in a fiber where we can neglect the linear coupling among the different modes, which is induced by random perturbations in the fiber structure such as manufacturing imperfections, environmental variations or local mechanism stress \cite{Palmieri14}.\\
For example, in a highly birefringent (HiBi) fiber the intrinsic large birefringence makes the propagation constants of all spatial and polarization modes to be noticeably different so as to nullify in practice the linear modal coupling.\\
On the contrary, in quasi-perfectly circular core fibers we find group of modes that are quasi-degenerate and may thus experience a strong linear coupling. Nevertheless Eq.(\ref{TotalEquation}) is still valid provided that the length $L$ of the fiber is small if compared to the characteristic linear coupling length of these modes. Actually, under this condition the fiber may be considered isotropic so that quasi-degenerate modes do not exchange energy and can thus be represented by an unique mode that is a proper linear combination of them.\\
We introduce here the nonlinear length $L_{NL}\equiv\big(\gamma T)^{-1}$, where $\gamma=max \{c_{klmn}\}$ is the largest among the nonlinear coupling coefficients and $T=\sum_n |P_{nx}|^2 + |P_{ny}|^2$ is the total input power.  This parameter provides an idea of the length scale of the nonlinear interactions and permits to select in Eq.(\ref{TotalEquation}) the only relevant nonlinear terms, that are those for which the condition $|\Delta\beta_{r,klmn} L_{NL}|\approx 0$ applies. Otherwise, if $|\Delta\beta_{r,klmn} L_{NL}|>> 0$, then the corresponding nonlinear term is rapidly oscillating and averages out to 0, so that it can be neglected. A proper discussion about the validity of this rule of thumb is done in Sec.~\ref{rif:limits}.\\
 In practice, due to the large phase-mismatch between the different spatial modes, in a single-core HiBi or isotropic fiber the condition $|\Delta\beta_{1,klmn} L_{NL}|\approx 0$  is achieved only if $\{k=m, l=n\}$ or $\{k=n, l=m\}$. Similarly $\{k=m, l=n\}$ or $\{k=l, m=n\}$ are needed in order to have $|\Delta\beta_{2,klmn} L_{NL}|\approx 0$. In a HiBi fiber, where a large phase-mismatch is present even between two polarization modes, the condition $|\Delta\beta_{3,klmn} L_{NL}|\approx 0$ cannot typically be attained, while $|\Delta\beta_{4,klmn} L_{NL}|\approx 0$ requires $\{l=k,m=n\}$.\\
Taking into account only the relevant nonlinear terms, Eq.(\ref{TotalEquation}) can be rewritten as follows in the case of an HiBi fiber:

 \begin{flalign}
&\frac{\partial {P_{nx}}}{\partial z} =- \it{v}_{nx}^{-1}\frac{\partial  P_{nx}}{\partial t}  - i\frac{\beta_{2nx}}{2}\frac{\partial^2  P_{nx}}{\partial t^2}+ \frac{\beta_{3nx}}{6}\frac{\partial^3  P_{nx}}{\partial t^3} - &&\nonumber\\
&\alpha_{nx}P_{nx} +i b_S C_{nn}|P_{nx}|^2P_{nx} + i\sum_{k\neq n} b_{||} C_{kn}|P_{kx}|^2 P_{nx}&&\nonumber\\
& +i b_X C_{nn}|P_{ny}|^2P_{nx} + i\sum_{k\neq n} b_{\bot} C_{kn}|P_{ky}|^2 P_{nx}
\label{TotalEquation2}
\end{flalign}

where $C_{kn}\equiv 3 c_{kknn}$; $b_S=1$ is the coefficient related to the self-phase modulation (spm); $b_X=2/3$ is the coefficient related to the intramodal cross-phase modulation (xpm); $b_{||}=2$ and $b_{\bot}=2/3$ are the coefficients related to the intermodal xpm involving modes with parallel or orthogonal polarization, respectively. \\
Equation (\ref{TotalEquation2}) holds also for an isotropic fiber in the scalar case, that is when the input field is linearly polarized along one fixed direction, let us say the $x$-axis (i.e. we set $b_X=b_{\bot}=0$). If the input field at the isotropic fiber is right or left circularly polarized, it proves convenient to rewrite Eq.(\ref{TotalEquation}) in terms of the circularly polarized components $P_{n+}\equiv (P_{nx}+i P_{ny})/\sqrt{2}$ and  $P_{n-}\equiv (P_{nx}-i P_{ny})/\sqrt{2}$, which brings to the same equality of Eq.(\ref{TotalEquation2}) after substitution of label $x$ ($y$) with $+$ ($-$) and setting $b_S=2/3$, $b_{||}=4/3$, $b_X=b_{\bot}=0$.\\
Differently from HiBi or short isotropic fibers, in telecommunication fibers rapid random variations of the fiber structure must be taken into account, as they lead to a non-negligible linear modal coupling. At this purpose, a generalized multimode Manakov model has been derived by Mumtaz et al. in \cite{Agrawal13}. It turns out that in the CW limit Eq.(\ref{TotalEquation2}) is still valid in a telecommunication fiber by setting $b_S=b_X=8/9$, $b_{||}=b_{\bot}=4/3$ and replacing $P_{nx},P_{ny}$ by the envelopes $\tilde{P}_{nx},\tilde{P}_{ny}$ that are obtained by means of a proper unitary transformation.\\
In the following we will indicate the field components as $P_{nx}$ ($P_{ny}$), with the assumption that they should read as $P_{n+}$ ($P_{n-}$) in the case of isotropic fibers where right/left circular polarization are involded, and as $\tilde{P}_{nx}$ ($\tilde{P}_{ny}$) in the case of telecommunication fibers.\\
As usual, in order to examine modulational instability processes, we introduce small amplitude perturbations. We thus decompose the modal envelope $P_{na}$ ($a=\{x,y\}$) in the sum of a pump $p$ and two perturbations $s$ and $i$ that indicate a signal Stokes and an idler anti-Stokes sideband symmetrically detuned with respect to the pump and centered at the frequencies $\omega_s$ and $\omega_i$, respectively:

 \begin{flalign}
 &P_{na}=p_{na}+s_{na}exp(i\Delta\beta_{na}^{(p,s)}z+i\Omega t)+&&\nonumber\\
&i_{na}exp(i\Delta\beta_{na}^{(p,i)}z-i\Omega t)
\label{Decomposition}
\end{flalign}

where $\Omega=\omega_s-\omega_p=\omega_p-\omega_i$ is the pump-sidebands detuning. The wavevector mismatches read as $\Delta\beta_{na}^{(p,s)}\equiv \beta_{s,na}-\beta_{p,na}$ and $\Delta\beta_{na}^{(p,i)}\equiv \beta_{i,na}-\beta_{p,na}$, where $\beta_{s,na}\equiv\beta_{na}(\omega_s)$ and $\beta_{i,na}\equiv\beta_{na}(\omega_i)$ indicate the propagation constant of the $na$-mode at the frequencies $\omega_s$ and $\omega_i$, respectively.\\
The decomposition Eq.(\ref{Decomposition}) is inserted in Eq.(\ref{TotalEquation2}) in order to find the set of differential equations that rules the evolution of pump and sidebands. \\
The equation for $p_{nx}$ is given by Eq.(\ref{TotalEquation2}) after substitution of P with p. We initially neglect propagation losses, namely we set $\alpha_{nx}=\alpha_{ny}=0$. The system dynamics in the presence of losses will be discussed later (see Appendix 2). Under this condition and in the CW-limit one obtains the following analytical solution:

 \begin{flalign}
&p_{nx}(z) = |p_{nx}|exp(i\psi_{nx}(0)+ i \phi_{nx}z )\nonumber\\
&\phi_{nx} =  b_S C_{nn}|p_{nx}|^2+b_X C_{nn}|p_{ny}|^2+&&\nonumber\\
&+\sum_{k\neq n} b_{||} C_{kn}|p_{kx}|^2  + \sum_{k\neq n} b_{\bot} C_{kn}|p_{ky}|^2
 \label{PumpSolution}
\end{flalign}

where $\psi_{nx}(0)$ is the phase of the input $p_{nx}(0)$. A similar analytical solution holds true for $p_{ny}$ after exchanging $x\leftrightarrow y$.\\
The equation for the modal envelope $s_{nx}$, obtained by linearization of Eq.(\ref{TotalEquation2}) and neglecting losses, reads as:

 \begin{flalign}
&\frac{\partial {s_{nx}}}{\partial z} =i(\phi_{nx}+ b_S C_{nn}|p_{nx}|^2)s_{nx} +&&\nonumber\\
&i b_X C_{nn} p_{ny}^*p_{nx}s_{ny}exp(i\Delta\beta_{ny}^{(p,s)}z-i\Delta\beta_{nx}^{(p,s)}z)+&&\nonumber\\
&i b_S C_{nn} p_{nx}^2 i_{nx}^*exp(-i\Delta\beta_{nx}^{(p,s)}z-i\Delta\beta_{nx}^{(p,i)}z)+&&\nonumber\\
&i b_XC_{nn} p_{nx}p_{ny} i_{ny}^*exp(-i\Delta\beta_{nx}^{(p,s)}z-i\Delta\beta_{ny}^{(p,i)}z)+&&\nonumber\\
&i \sum_{k\neq n} b_{||} C_{kn} p_{kx}^*p_{nx} s_{kx} exp(i\Delta\beta_{kx}^{(p,s)}z-i\Delta\beta_{nx}^{(p,s)}z)+&&\nonumber\\
&i \sum_{k\neq n} b_{\bot} C_{kn} p_{ky}^*p_{nx} s_{ky} exp(i\Delta\beta_{ky}^{(p,s)}z-i\Delta\beta_{nx}^{(p,s)}z)+&&\nonumber\\
&i \sum_{k\neq n} b_{||} C_{kn} p_{kx} p_{nx} i_{kx}^* exp(-i\Delta\beta_{kx}^{(p,i)}z-i\Delta\beta_{nx}^{(p,s)}z)+&&\nonumber\\
&i \sum_{k\neq n}b_{\bot} C_{kn} p_{ky} p_{nx} i_{ky}^*exp(-i\Delta\beta_{ky}^{(p,i)}z-i\Delta\beta_{nx}^{(p,s)}z)
\label{SidebandEquation}
\end{flalign}

Similar equations apply to $s_{ny}$ after exchanging $x\leftrightarrow y$, to $i_{nx}$ after exchanging $i\leftrightarrow s$ and to $i_{ny}$ after exchanging $x\leftrightarrow y$, $i\leftrightarrow s$.\\
In order to get rid of the oscillating terms, it proves useful to employ the following change of variables:

\begin{flalign}
&s_{na}(z) = \bar{s}_{na}(z)exp( i\psi_{na}(0) + i \phi_{na}z -i\Delta\beta_{na}^{(p,s)}z )&&\nonumber\\
&i_{na}(z) = \bar{i}_{na}(z)exp( i\psi_{na}(0) + i \phi_{na}z -i\Delta\beta_{na}^{(p,i)}z )&&\nonumber\\
\label{Transformations}
\end{flalign}

After insertion of Eq.(\ref{Transformations}) in Eq.(\ref{SidebandEquation}) and rewriting the fields $p_{nx},p_{ny}$ according to the solution given in Eq.(\ref{PumpSolution}), we finally obtain the following eigenvalue problem:

\begin{flalign}
&\partial_z{\bf v} = i{\bf M}{\bf v}&&\nonumber\\
&{\bf v}=[{\bf s_x}\,\,\, {\bf s_y }\,\,\,{\bf i_x}^* \,\,\, {\bf i_y}^*]^T &&\nonumber\\
&{\bf M } =
\begin{bmatrix}  \bf{M_{sx,sx}}& \bf{M_{sx,sy}} & \bf{M_{sx,ix}}& \bf{M_{sx,iy}}\\
                         \bf{M_{sy,sx}}& \bf{M_{sy,sy}} & \bf{M_{sy,ix}}& \bf{M_{sy,iy}}\\
                         -\bf{M_{ix,sx}}& -\bf{M_{ix,sy}} & -\bf{M_{ix,ix}}& -\bf{M_{ix,iy}}\\
                         -\bf{M_{iy,sx}}& -\bf{M_{iy,sy}} & -\bf{M_{iy,ix}}& -\bf{M_{iy,iy}}
\end{bmatrix}
\label{LinearSystem}
\end{flalign}

where ${\bf s_a}=[\bar{s}_{1a}\,\, \bar{s}_{2a}\,\, ...\bar{s}_{Na}]^T$ and ${\bf i_a}=[\bar{i}_{1a}\,\, \bar{i}_{2a}\,\, ...\bar{i}_{Na}]^T$  ($a=\{x,y\}$) are $N\times1$ vectors, and ${\bf M}$ is a $4N\times4N$ matrix composed by 16 $N\times N$ matrix blocks.\\
Each matrix block accounts for the interaction between different subsets of sideband modes. For example, the block $\bf {M_{sx,sx}}$ is related to the mutual interaction between the $x-$polarized signal modes, whereas $\bf {M_{sx,sy}}$ is related to the interaction between the $x-$ and $y-$polarized signal modes. The elements of ${\bf M_{sx,sx}}$,  ${\bf M_{sx,sy}}$ and ${\bf M_{sx,ix}}$ are:

\begin{flalign}
& {\bf M_{sx,sx} }[n,n]= \Delta\beta_{nx}^{(p,s)}+ b_S C_{nn}|p_{nx}|^2&&\nonumber\\
&{\bf M_{sx,sx} }[n,m] = b_{||} C_{mn}|p_{mx}||p_{nx}|\,\,\,(n \neq m)&&\nonumber\\
& {\bf M_{sx,sy} }[n,n] = b_X C_{nn} |p_{ny}||p_{nx}|&&\nonumber\\
&{\bf M_{sx,sy} }[n,m]= b_{\bot} C_{mn} |p_{my}||p_{nx}|\,\,\,(n \neq m)&&\nonumber\\
& {\bf M_{sx,ix} }[n,n]= b_S C_{nn}|p_{nx}|^2&&\nonumber\\
&{\bf M_{sx,ix} }[n,m]= b_{||} C_{mn} |p_{mx}||p_{nx}|\,\,\,(n \neq m)&&\nonumber\\
\label{MatrixBlocks}
\end{flalign}

Furthermore ${\bf M_{sx,iy} }={\bf M_{sy,ix} }^T={\bf M_{ix,iy} }={\bf M_{sx,sy} }$; the matrix ${\bf M_{sy,sy} }$  reads as ${\bf M_{sx,sx} }$ after replacing $x$ with $y$; ${\bf M_{ix,ix} }$ reads as ${\bf M_{sx,sx} }$ after replacing $s$ with $i$; ${\bf M_{iy,iy} }$ reads as ${\bf M_{sy,sy} }$ after replacing $s$ with $i$; ${\bf M_{sy,iy} }$ reads as ${\bf M_{sx,ix} }$ after replacing $x$ with $y$; ${\bf M_{sy,sx} }={\bf M_{sx,sy} }^T$; ${\bf M_{ix,sx} }={\bf M_{sx,ix} }$; ${\bf M_{ix,sy} }={\bf M_{sy,ix} }^T$; ${\bf M_{sy,sx} }={\bf M_{sx,sy} }^T$; ${\bf M_{iy,sx} }={\bf M_{sx,iy} }^T$ ; ${\bf M_{iy,sy} }={\bf M_{sy,iy} }$; ${\bf M_{iy,ix} }={\bf M_{ix,iy} }^T$.\\
The solution of Eq.(\ref{LinearSystem}) reads as:

\begin{flalign}
v[j](z) = \sum_{k=1}^{4N} c_k w_k[j] exp(i\lambda_k z)
\label{EvolutionV}
\end{flalign}

where  $\lambda_k$ and ${\bf w}_k$ ($1\leq k \leq 4N$) are respectively the eigenvalues and the eigenvectors of ${\bf M}$, whereas $v[j]$ ($w_k[j]$) indicates the $j-$element of ${\bf v}$ (${\bf w}_k$). The coefficient $c_k$ is given by the projection of the input ${\bf v}(z=0)$, which is fixed by the input modal envelopes of the sidebands, over the set of eigenvectors. We highlight that the structure of the matrix $\bf{M}$ guarantees that its eigenvectors satisfy the orthogonality relation ${\bf w}_{k1}\bullet{\bf D} {\bf w}_{k2}=0$ $(k_1 \neq k_2)$, where ${\bf D}=diag([ones_{2N}, \,\,-ones_{2N}])$ is a diagonal matrix, $ones_{2N}$ is a $1 \times 2N$ vector of ones and $\bullet$ indicates the scalar product. By exploiting this orthogonality relation we can easily compute $c_k$ as follows: $c_k =( {\bf v}(z=0)\bullet{\bf D}{\bf w_k} )/ ( {\bf w_k}\bullet{\bf D}{\bf w_k} )$.\\
An eigenvalue $\lambda_k$ with negative imaginary part leads to an exponential amplification of $v[j]$ which is characterized by a gain $g_k \equiv -Im(\lambda_k)$. We may be tempted to approximate Eq.(\ref{EvolutionV}) by taking into account only the largest gain $g_{max}$ and its corresponding eigenvector ${\bf w_{max}}$, as they ultimately determine the asymptotic  evolution of $v[j]$, namely $|{v}[j](z)| \approx |c_{max}||{w} _{max}[j]| exp(g_{max}z)$.\\
Nevertheless, in doing so, we may not properly describe the evolution of ${v}[j]$ when two or more eigenvalues with negative imaginary part are present. Indeed two gains $g_1$ and $g_2$ could exist such that $g_2>g_1$ but $|c_1 w_1[j]| exp(g_1 \tilde{z})>|c_2 w_2[j]| exp(g_2 \tilde{z})$ at a certain position $\tilde{z}$. In this case we will define $g_1$  as the {\it dominant gain} for ${v}[j]$ at the position $\tilde{z}$, as it plays the role of true leading term in the MI growth of $v[j]$.\\
From what discussed above, a good approximation of $|v[j]|$ at the fiber exit, let us say in z=L, could be obtained by computing the corresponding dominant gain, that is $|v[j]|(L)\approx max_k \{|c_k w_k[j]| exp(g_k L)\}$. The signal and idler modal amplitudes at the fiber exit can be finally estimated by noting that $|s_{nx}|\equiv |v[n]|$; $|s_{ny}|\equiv|v[n+N]|$; $|i_{nx}|\equiv|v[n+2N]|$; $|i_{ny}|\equiv|v[n+3N]|$ ($1\leq n \leq N$).\\
We underline that the coupling coefficients $C_{kn}$ of the matrix $\bf{M}$ are generally $\Omega$-dependent, and so are the wavevector mismatches $\Delta\beta_{na}^{(p,s)}$ and $\Delta\beta_{na}^{(p,i)}$. Both the mismatches could be approximated by expanding $\beta_{na}(\omega_s)$ and $\beta_{na}(\omega_i)$ in a Taylor series centered around $\beta_{na}(\omega_p)$. Truncating the expansion at the third-order we finally get:

\begin{flalign}
&\Delta\beta_{na}^{(p,s)}=  \Omega/v_{na} + \Omega^2\beta_{2na}/2 + \Omega^3\beta_{3na}/6\nonumber\\
&\Delta\beta_{na}^{(p,i)}= -\Omega/v_{na} + \Omega^2\beta_{2na}/2 - \Omega^3\beta_{3na}/6
\label{TaylorExpansion}
\end{flalign}

Therefore for a given detuning $\Omega$ and knowing the input modal envelopes of both pump and sidebands we may compute the corresponding eigenvalues and eigenvectors of ${\bf M}$ and then estimate the output modal envelopes $s_{na}(L)$ and $i_{na}(L)$ as formerly discussed.\\
From what stated it clearly appears that the MI growth of the sidebands depends on the eigenvectors and eigenvalues of ${\bf M}$, therefore it can be controlled by means of the system parameters, such as the modal power distribution of the pump, the nonlinear coupling coefficients and the modal dispersion characteristics (that is group velocity, GVD and TOD of each mode).\\
This outcome opens the way towards parametric amplification in multimode fibers and represents the core of the IM-MI analysis developed in this paper. It generalizes previous results on vectorial MI in single-mode fibers, for which the matrix ${\bf M}$ is $4 \times 4$, to the most general case of the IM-MI in a fiber where N spatial and polarization modes propagate. Furthermore, as previously observed, the theory developed in this section applies to all the main types of commonly used optical fibers.\\

\section{Amplification of an input ASE noise}\label{rif:simulations}
\begin{figure}[htbp]
   \begin{center}
       \includegraphics[width=0.85\columnwidth]{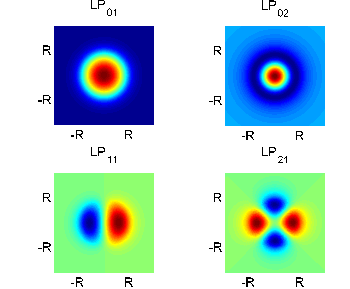}
   \end{center}
   \caption{Modal transverse profile of the 4 modes supported by the simulated fiber. The radius of the core is $R=12\mu m$.}
   \label{figModes}
\end{figure}

In order to provide numerical evidence confirming the theoretical results previously argued, we simulate the propagation in a multimode fiber by numerically solving Eq.(\ref{TotalEquation2}) with the split-step Fourier method. Finally, we compare the output modal amplitudes obtained from numerical simulations with those inferred from the analytical outcomes discussed in Section ~\ref{rif:general}.\\
With the aim of focusing on the most striking features of the IM-MI, we represent a simple but effective example: the propagation in an isotropic fiber of 4 spatial modes that are linearly polarized along the same direction.\\ 
The simulated fiber is circular in cross-section and step-index, which allows easily solving the well-known modal characteristic equation in order to compute the modal transverse profiles and the main modal parameters (Fig.~\ref{figModes} and Table 1-2).\\
The CW input field is centered at the carrier wavelength $\lambda=1550$ nm and is linearly polarized along the $x-$direction. The diameter of the core is $2R=24\mu m$, the refractive indexes of cladding and core are $n_{clad}=1.5$ and $n_{core}=1.5035$, respectively.  The corresponding V-number of the fiber $V\equiv 2\pi R \lambda^{-1}(n_{core}^2-n_{clad}^2)^{1/2}$ is $V\approx5$, so that 4 non-degenerate spatial modes propagates : $LP_{01}$ , $LP_{02}$, $LP_{11}$ and $LP_{21}$. In our case these modes are $x-$polarized and will be indicated respectively as $1x-$mode, $2x-$mode, $3x-$mode and $4x-$mode.\\
We don't take into account for the 2-fold degeneracy of the $LP_{11}$ mode: indeed we assume that only one among the two degenerate modes is excited at the input fiber and that it does not couple to the second because the fiber is ideally isotropic. A similar consideration applies to the $LP_{21}$ mode.

\begin{table}[h]
\caption{Propagation constant $\beta$, group velocity mismatch $(GVM)$, group velocity dispersion $\beta_2$  and 3rd order dispersion $\beta_3$ of the 4 supported modes at the pump wavelength 1550nm.These parameters have been adjusted in order to limit the IM-MI bandwidth to nearly 30 THz, so to avoid higher-order dispersion terms besides the TOD. Note that in a real fiber they can be adjusted over a wide range of values by means of an appropriate fiber design.}
\begin{tabular}{|l|l|l|l|l|}
\hline
                                        & $\bf{LP_{01}}$ & $\bf{LP_{02}}$ & $\bf{LP_{11}}$ & $\bf{LP_{21}}$      \\ \hline
$\beta [\mu m^{-1}]$      & 6.0995         & 6.0836         & 6.0891         & 6.0848             \\ \hline
$GVM[ps m^{-1}]$       & 0         & 10.8         & 7.1         & 13.6             \\ \hline
$\beta_2[ps^2 km^{-1}]$   &  21.7           &   -147.7          &    36.3        &       -3.5\\ \hline
$\beta_3[fs^3 mm^{-1}]$   &  89.5           &    -7361.1       &    -169.9        & -2128.7       \\ \hline
\end{tabular}
\end{table}

\begin{table}[h]
\caption{ Nonlinear coupling coefficients $C_{kn}$ of the simulated fiber. The coefficients are normalized with respect to $C_{11}=10 W^{-1}km^{-1}$, so that the element in row $k$ and column $n$ represents $C_{kn}/C_{11}$.}
\begin{tabular}{|l|l|l|l|l|}
\hline
 &          $\bf{n=1}$ & $\bf{n=2}$ & $\bf{n=3}$ & $\bf{n=4}$ \\ \hline
 $\bf{k=1}$   &1.00      &0.73    &0.66    &0.45\\ \hline
 $\bf{k=2}$   &0.73      &0.96    &0.37    &0.33\\ \hline
 $\bf{k=3}$   &0.66      &0.37    &1.04    &0.61\\ \hline
 $\bf{k=4}$   &0.45      &0.33    &0.61    &0.92\\ \hline
\end{tabular}
\end{table}


We set the total input power equal to $4000 W$ and uniformly distributed over the 4 pump modes, therefore $|p_{nx}|^2=1000 W$ (n=\{1,2,3,4\}). { \bf Such an high-power injection is indeed feasible in multimode fibers when employing nanosecond pump sources at low rates, so that the pump can be practically considered CW \cite{Millot97,Modotto11}.}
We also add a weak background white noise, which could be the amplified spontaneous emission (ASE) noise in a realistic experiment. The total input field for the $nx-$mode is thus written as $P_{nx}(0,t)= p_{nx}(0) + r_{nx}(0,t)$, being $p_{nx}(0)=(1000 W)^{1/2}$ and $r_{nx}(0,t)$ a white noise generated by adding at each frequency component a random variable with independent Gaussian-distributed real and imaginary parts.\\
We finally solve Eq.(\ref{TotalEquation2}) by split-step Fourier method using $P_{nx}(0,t)$ as input field for the $nx-$mode.\\
Each pair of noise frequency samples centered at $+\Omega$ and $-\Omega$ play the role of small input signal and idler perturbations that are amplified by the IM-MI process. The power spectrum $\hat{R}_{nx}(L,\Omega)$ of $r_{nx}(L,t)$ reveals the amount of power carried by the $nx-$mode at the detuning $\Omega$ and at the fiber exit $z=L$. As explained in Appendix 1, we make use of the averaged power spectrum $\hat{R}_{nx,avg}(L,\Omega)$ (Fig.~\ref{figAppendix}), which expresses the truly noise power level around $\Omega$.\\
The function $\hat{R}_{nx,avg}(L,\Omega)/\hat{R}_{nx,avg}(0,\Omega)$ defines the ratio between the output and the input power coupled to the $nx$-mode, and the logarithmic square-rooted power-ratio normalized to distance, here indicated with $\hat{A}_{nx}(L,\Omega)$, represents the amplitude amplification factor for the $nx-$mode at the fiber output:

\begin{flalign}
\hat{A}_{nx}(L,\Omega)=\frac{1}{L}Log \left(\frac{\hat{R}_{nx,avg}(L,\Omega)^{1/2}}{\hat{R}_{nx,avg}(0,\Omega)^{1/2}}\right)
\label{AmplificationFactor}
\end{flalign}

We will compute the function $\hat{A}_{nx}$ obtained from numerical solution of Eq.(\ref{TotalEquation2}) and we will compare it with the following analytical estimate $\hat{A}_{nx,est}$ gathered from the results exposed in the previous section (see Appendix 1):


 \begin{flalign}
&\hat{A}_{nx,est}(L,\Omega)= g_{dom,nx} + L^{-1}Log(|\tilde{w}_{dom,nx}[n]|)\nonumber\\
&{\bf\tilde{w}_{dom,nx}}={\bf w_{dom,nx}}/({\bf w_{dom,nx}}\bullet{\bf D}{\bf w_{dom,nx}})\nonumber\\
& g_{dom,nx}\equiv g_{dom,nx}(L,\Omega)\,\,\,\,\,\,\, {\bf w_{dom,nx}} \equiv {\bf w_{dom,nx}}(L,\Omega)
 \label{SpectralEstimation}
 \end{flalign}

where $g_{dom,nx}$ is the dominant gain for the $nx-$mode at the detuning $\Omega$ and at the position $z=L$, whereas $\bf{w}_{dom,nx}$ is its corresponding eigenvector, and they satisfy the following condition:

 \begin{flalign}
 |\tilde{w}_{dom,nx}[n]| exp\big(g_{dom,nx} L\big)=max_k\big\{ |\tilde{w}_k[n]| exp(g_k L) \big\}
 \label{DominantCondition}
 \end{flalign}

\begin{figure}[htbp]
   \begin{center}
       \includegraphics[width=0.9\columnwidth]{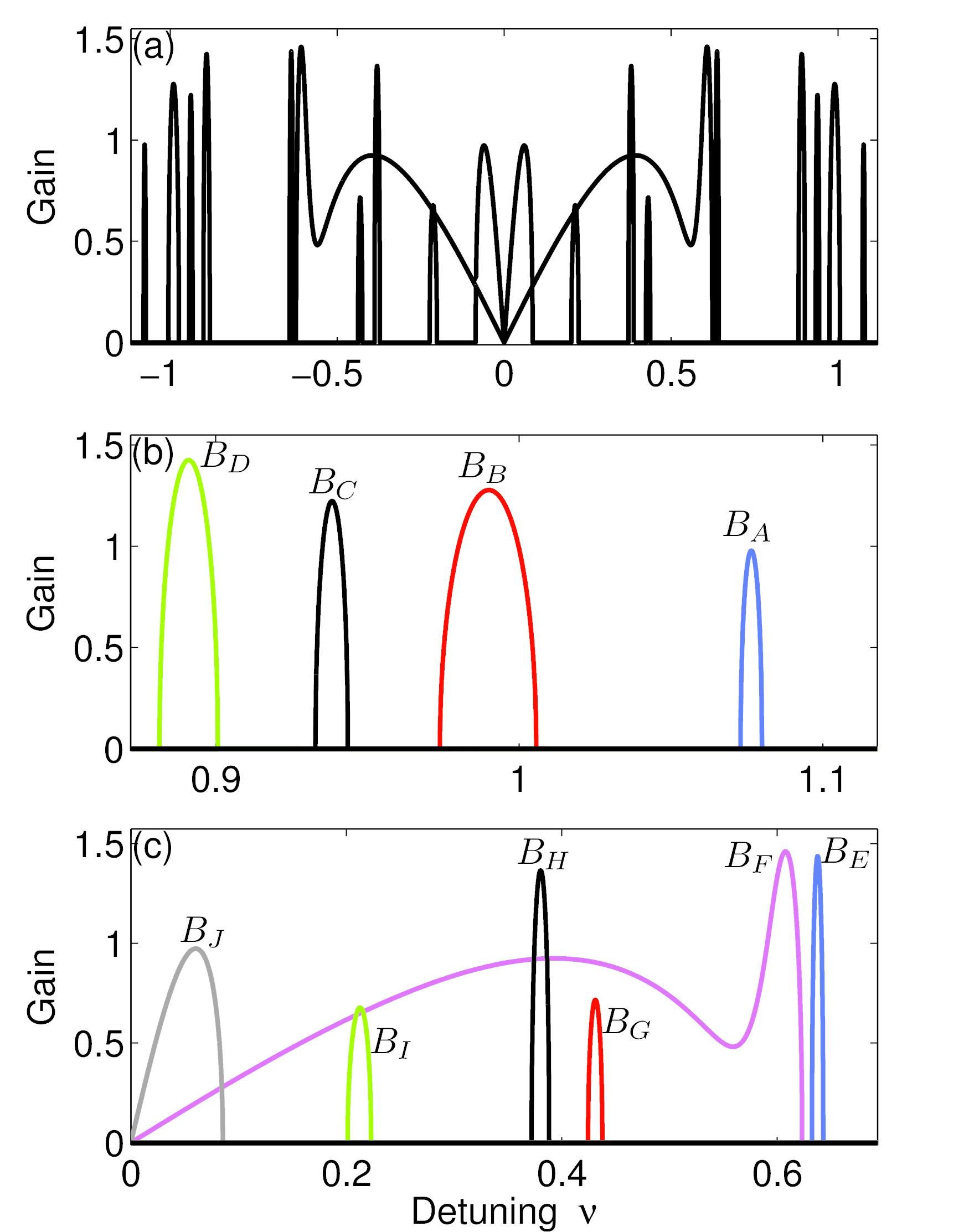}
   \end{center}
   \caption{normalized IM-MI gain VS normalized frequency detuning $\nu$. Gain values are normalized with respect to the nonlinear length. Different gain curves are found which are plotted with different colors and labeled with $B_A, B_B,...,B_J$. Panel (a): global view. Panel (b): zoom in the band from $\nu=0.85$ to $\nu=1.1$.Panel (c): zoom in the band from $\nu=0$ to $\nu=0.65$. Note that the IM-MI gain is symmetric with respect to $\nu=0$, therefore the same gain curves are found at positive and negative detunings. }
   \label{figGains}
\end{figure}

 Note that the third line of Eq.(\ref{SpectralEstimation}) has been added to emphasize the dependence of $g_{dom,nx}$ and ${\bf w_{dom,nx}}$ on both $L$ and $\Omega$ and that Eq.(\ref{SpectralEstimation}) is valid whether $\hat{A}_{nx,est}(L,\Omega)>0$, otherwise we set it equal to 0.\\
In order to find $g_{dom,nx}$ and $\bf w_{dom,nx}$  we build the $16\times 16$ matrix ${\bf M}$ of Eq.(\ref{LinearSystem}) by setting $|p_{nx}|=(1000 W)^{1/2}$, $|p_{ny}|=0 $ (n=\{1,2,3,4\}) and using the modal parameters and coupling coefficients displayed in Table 1-2. At each detuning $\Omega$ we then compute the corresponding gain coefficients $ g_k\equiv -\Im(\lambda_k)$ and eigenvectors ${\bf w_k}$ ($1 \leq k \leq 16$) and we look for the couple $(g_k,{\bf w_k})$ that fulfill Eq.(\ref{DominantCondition}).\\
We point out that the structure of the matrix ${\bf M}$ guarantees that $g_k(\Omega)=g_k(-\Omega)$. Furthermore the idler component $w_k[n+2N]$ is directly tied to the signal component $w_k[n]$ by the relation $w_k[n+2N](\Omega)=w_k[n](-\Omega)$. For this reason we compute only $w_k[n]$ as function of $\Omega$, assuming that positive frequencies refer to the signal $nx-$mode, whereas negative frequencies to the idler $nx-$mode. Note also that an equation similar to Eq.(\ref{SpectralEstimation}) would apply for the amplification factor $\hat{A}_{ny,est}(L,\Omega)$, here not considered as the modes are x-polarized, after replacing of $x$ with $y$ and of $w_k[n]$ with $w_k[n+N]$.  \\
For convenience in the following we make use of the normalized dimensionless distance $\xi\equiv z/L_{NL,1}$ and detuning $\nu\equiv(2\pi)^{-1}\Omega T_{NL,1}$, where $L_{NL,1}$ is the nonlinear length related to the $1x-$mode and $T_{NL,1}=(|\beta_{2,1x}|L_{NL,1}/2)^{1/2}$ is the corresponding characteristic nonlinear time. According to this normalization, in the example under discussion $\xi=1$ corresponds to $z=0.1 m$ and $\nu=1$ corresponds to a detuning $(2\pi)^{-1}\Omega=30.1 THz$, whereas a normalized gain $g=1$ corresponds to a real gain of $10m^{-1}$.\\
In Fig.~\ref{figGains} the gain coefficients are displayed as function of $\nu$.\\
We notice the existence of several gain curves, which is due to the rich set of modal interactions. This point will be treated in detail in the next section.

\begin{figure}[htbp]
   \begin{center}
       \includegraphics[width=0.9\columnwidth]{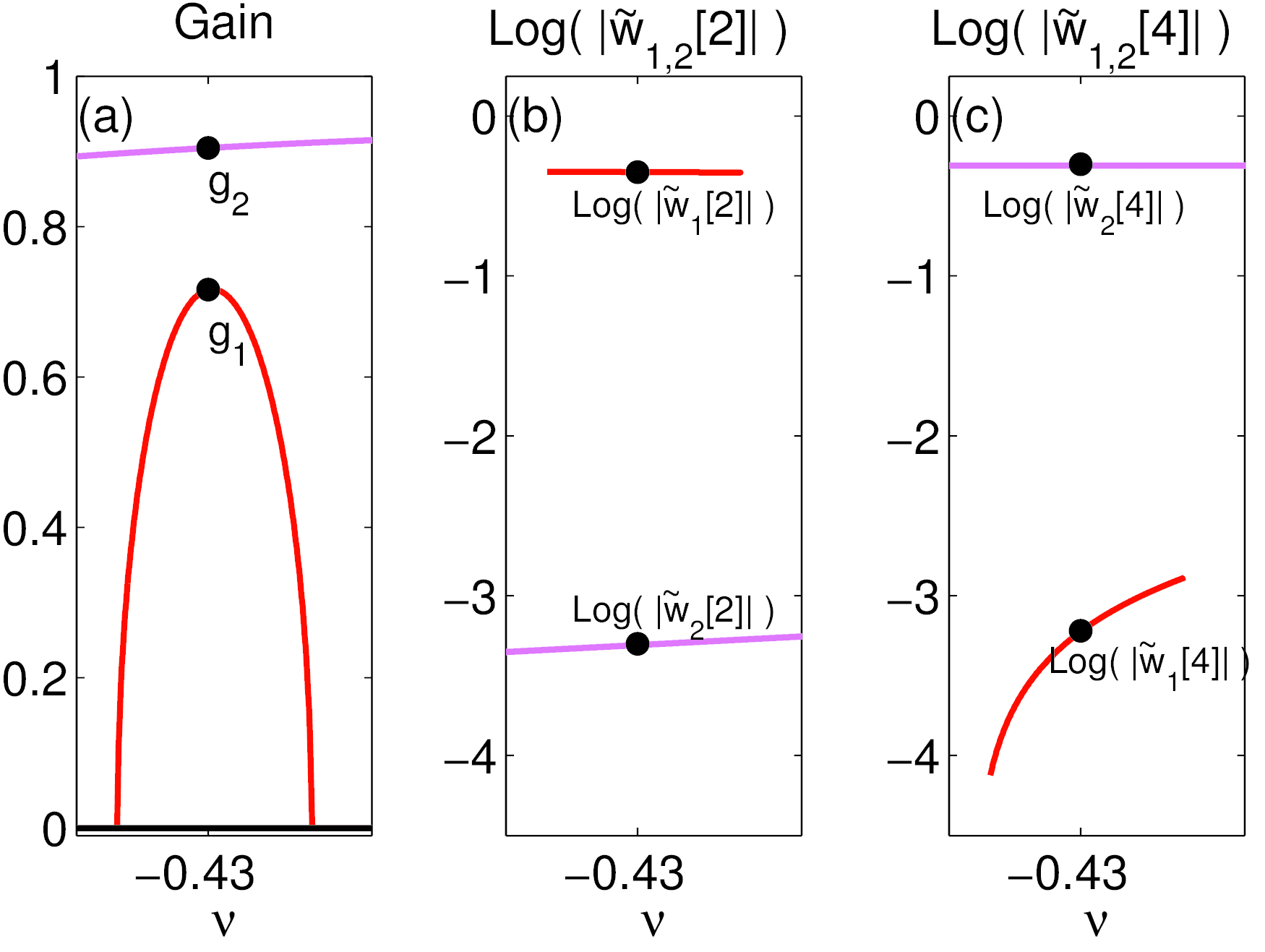}
   \end{center}
   \caption{Panel(a): IM-MI gain around $\nu=-0.43$. We recognize the gain curves $B_G$ (red) and $B_F$ (magenta). The black dots identifies $g_1\equiv B_G(\nu=-0.43)=0.71$ and $g_2\equiv B_F(\nu=-0.43)=0.9$. Panel(b): eigenvector components related to the $2x$-mode:  $Log(|\tilde{w}_1[2]|)$ (red) and  $Log(|\tilde{w}_2[2]|)$ (magenta).The black dots identifies $Log(|\tilde{w}_1[2]|)=-0.35$ and $Log(|\tilde{w}_2[2]|)=-3.35$ computed at the detuning $\nu=-0.43$. Panel(c): eigenvector components related to the $4x$-mode:  $Log(|\tilde{w}_1[4]|)$ (red) and  $Log(|\tilde{w}_2[4]|)$ (magenta). The black dots identifies $Log(|\tilde{w}_1[4]|)=-3.22$ and $Log(|\tilde{w}_2[4]|)=-0.35$ computed at the detuning $\nu=-0.43$.}
   \label{figEgv}
\end{figure}

\begin{figure}[htbp]
   \begin{center}
       \includegraphics[width=0.9\columnwidth]{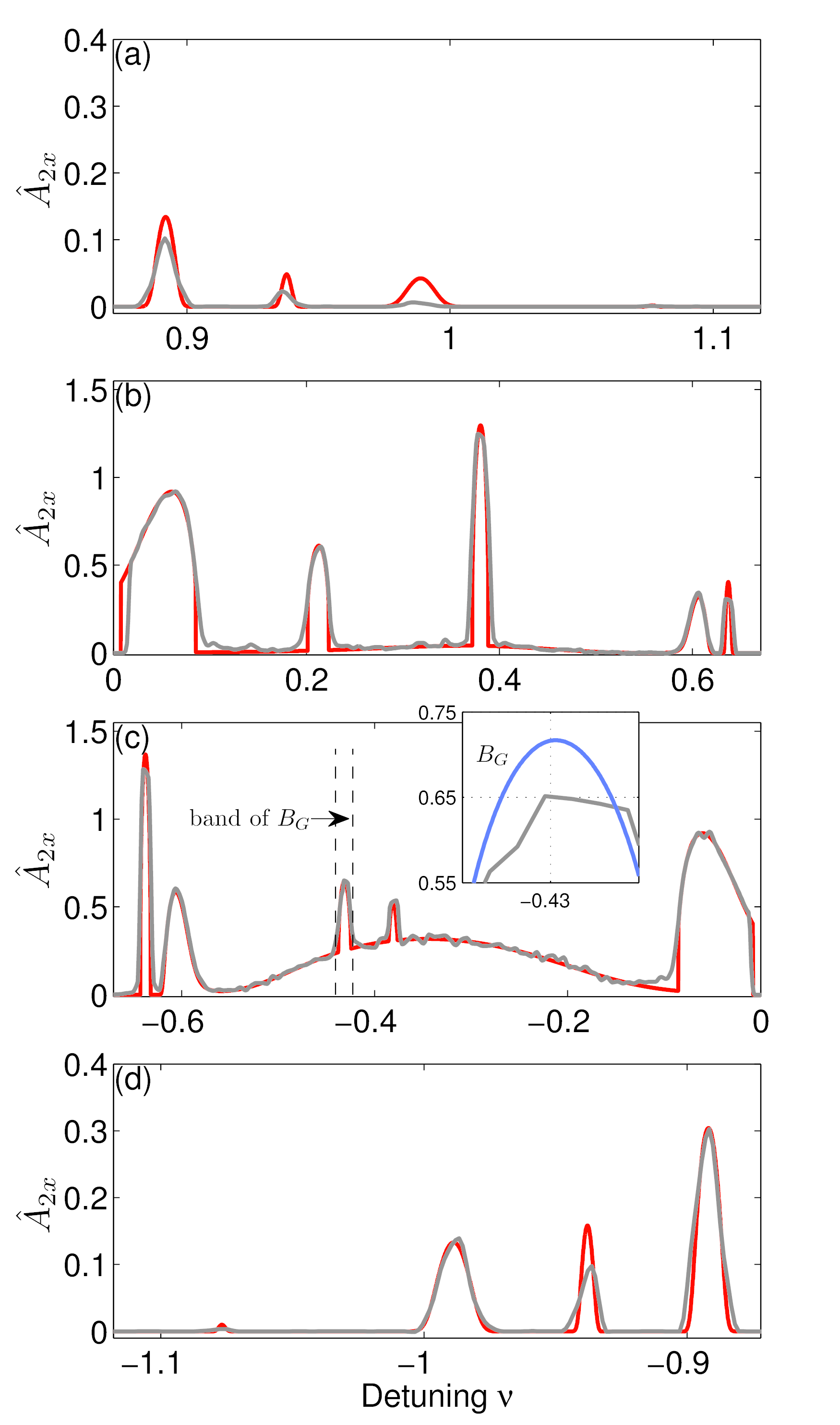}
   \end{center}
   \caption{Amplitude amplification factor for the $2x-$mode when $L=5$. Frequencies around $\nu=0$ (where the pump is located) have been filtered out. Panel (a): band from $\nu=0.85$ to $\nu=1.1$; (b): band from $\nu=0$ to $\nu=0.65$; (c): band from $\nu=-0.65$ to $\nu=0$; (d) band from $\nu=-1.1$ to $\nu=-0.85$. In gray: $\hat{A}_{2x}$ obtained by numerical solution of Eq.(\ref{TotalEquation2}) by split-step method. In red: the estimation $\hat{A}_{2x,est}$ obtained by Eq.(\ref{SpectralEstimation}). The vertical dashed lines in panel (c) delimit the band of $B_G$ centered at $\nu=-0.43$. The inset in panel (c) shows a zoom of $\hat{A}_{2x}$ around $\nu=-0.43$; the gain curve $B_G$ (blue) is also reported. The difference between the peaks of the two curves is nearly 0.07, that is to say $\hat{A}_{2x}\approx B_G-0.07$ in the band of $B_G$.}
\label{figNumericsMode2a}
\end{figure}

\begin{figure}[htbp]
   \begin{center}
       \includegraphics[width=0.9\columnwidth]{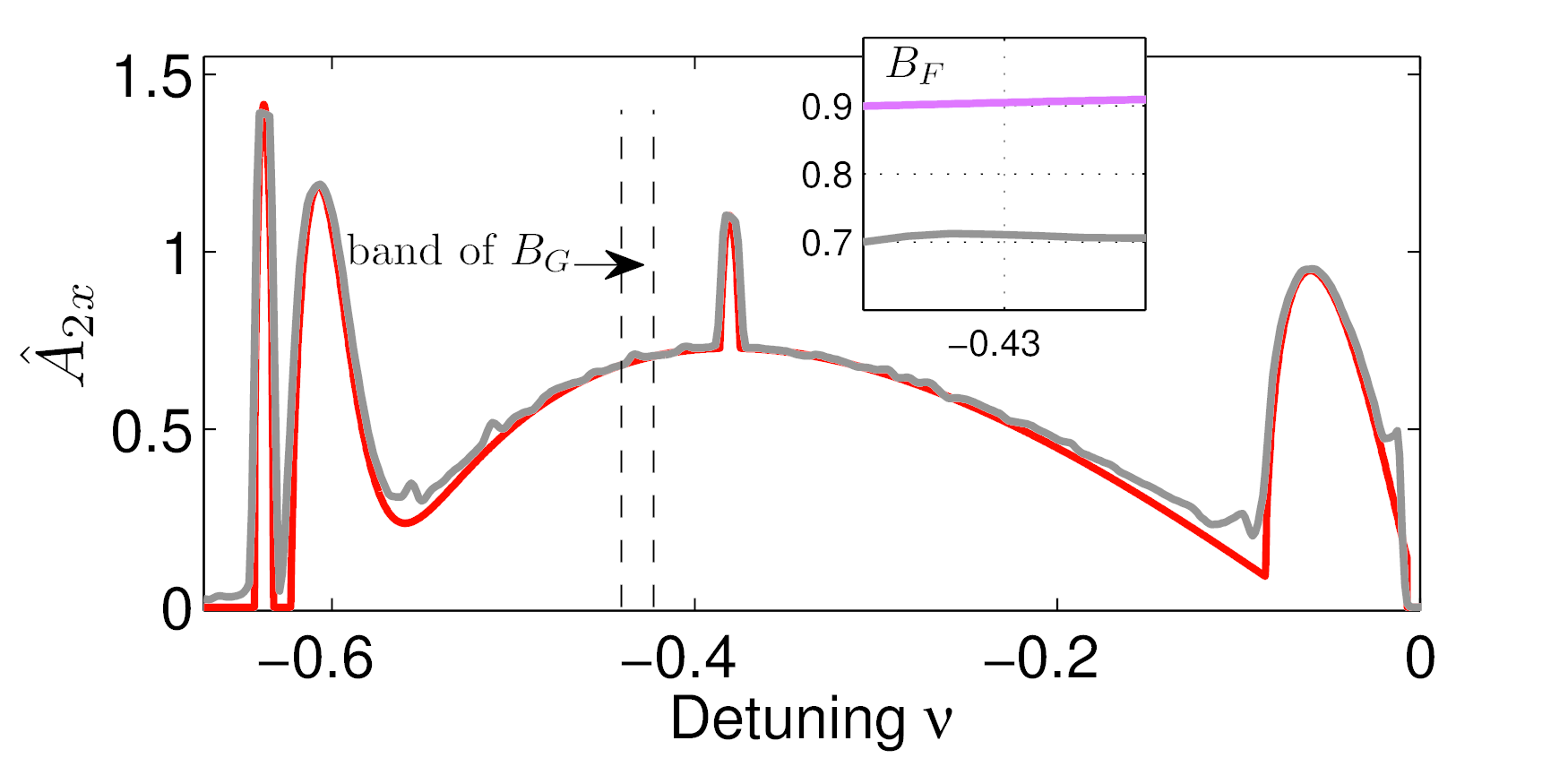}
   \end{center}
   \caption{Same as in Fig.~\ref{figNumericsMode2a}(c), but when $L=16$. The inset shows a zoom of $\hat{A}_{2x}$ around $\nu=-0.43$; the gain curve $B_F$ (magenta) is also reported. The difference between the two functions is nearly 0.21, that is to say $\hat{A}_{2x}\approx B_F-0.21$ in the band of $B_G$.}
   \label{figNumericsMode2b}
\end{figure}

\begin{figure}[htbp]
   \begin{center}
       \includegraphics[width=0.9\columnwidth]{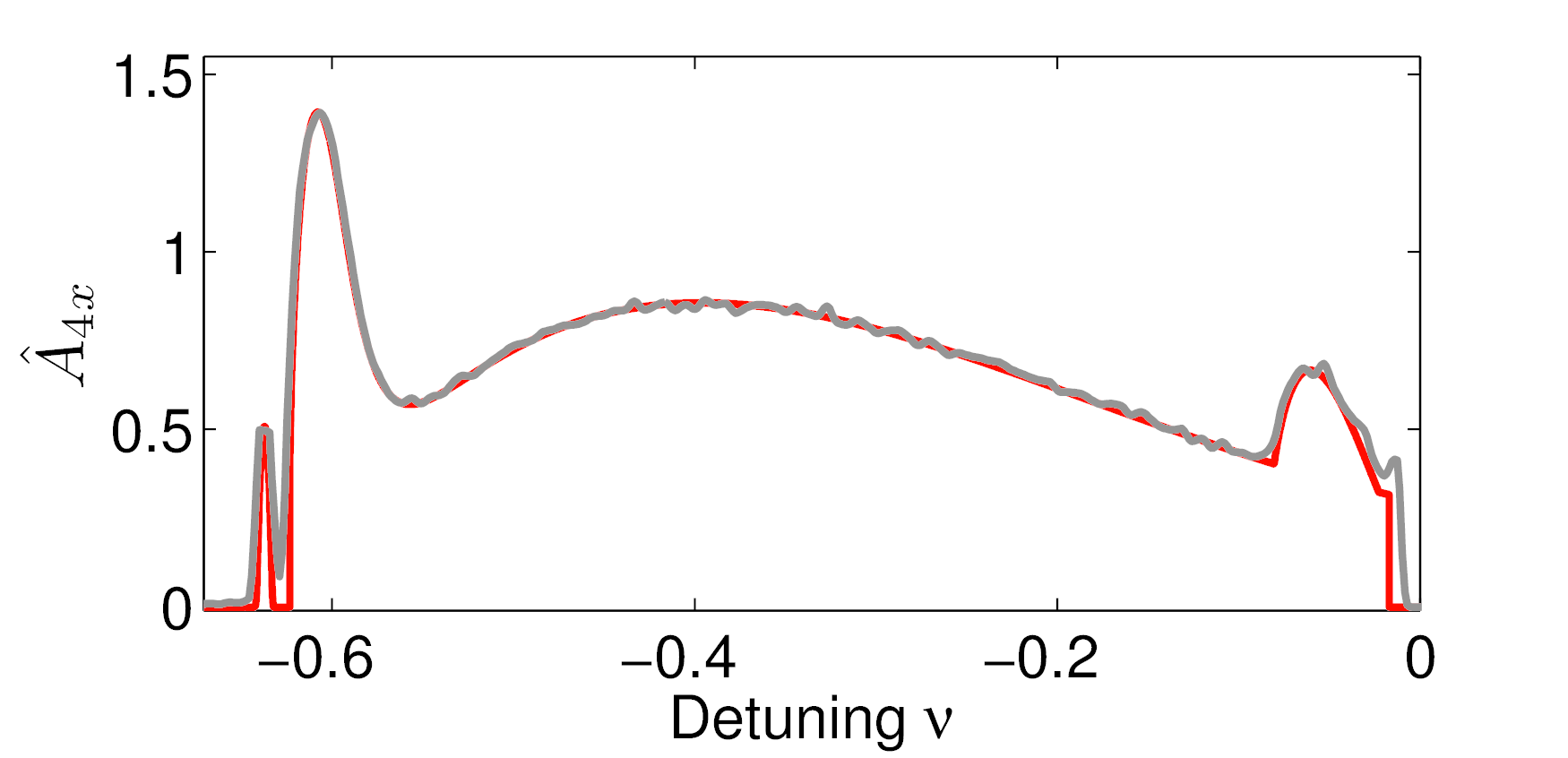}
   \end{center}
\caption{Amplitude amplification factor for the $4x-$mode when $L=5$ and in the band from $\nu=-0.65$ to $\nu=0$.Frequencies around $\nu=0$ (where the pump is located) have been filtered out. In gray: $\hat{A}_{4x}$ obtained by numerical solution of Eq.(\ref{TotalEquation2}) by split-step method. In red: the estimation $\hat{A}_{4x,est}$ obtained by Eq.(\ref{SpectralEstimation}). }
   \label{figNumericsMode4}
\end{figure}

 Here we highlight that some of the gain curves overlap, that is to say, for some value of $\nu$ two gains coexist that are dominant at different fiber positions. Without any loss of generality, we assume that the two gains come from the first two eigenvalues of ${\bf M}$: therefore we indicate with $g_1\equiv -\Im(\lambda_1)$ and $g_2\equiv-\Im(\lambda_2)$ the concurrent gains and with $\bf{w_1}$ and $\bf{w_2}$ the corresponding eigenvectors .\\
According to Eq.(\ref{DominantCondition}) and assuming $g_2>g_1$, if $|\tilde{w}_1[n]|>|\tilde{w}_2[n]|$  then the gain $g_1$ is dominant, for the $nx-$mode, whenever $ |\tilde{w}_1[n]| exp(g_1 z)>|\tilde{w}_2[n]| exp(g_2 z)$, that is when $z < (l_1 - l_2 )/(g_2-g_1)$, being $l_1\equiv Log(|\tilde{w}_1[n]|)$ and $l_2\equiv Log(|\tilde{w}_2[n]|)$. Otherwise, if $|\tilde{w}_1[n]|<|\tilde{w}_2[n]|$, then $g_2$ is dominant at any fiber position  for the $nx-$mode.\\
As example, let us consider the gain curves $B_F$ and $B_G$, which overlap in the whole band of $B_G$ (see Fig.~\ref{figGains}). Figure ~\ref{figEgv}a shows a zoom of $B_F$ and $B_G$ around the detuning $\nu=-0.43$. We observe that for $\nu=-0.43$ the peak $g_1=0.71$ of $B_G$ is lower than the gain value $g_2=0.90$ of $B_F$; nevertheless $Log(|\tilde{w}_1[2]|)\approx-0.35$ is larger than $Log(|\tilde{w}_2[2]|)\approx-3.35$ (Fig.~\ref{figEgv}b), which makes $g_1$ to be dominant for the $2x-$mode until $\xi=15.8$. More generally we can conclude that for any detuning close to $\nu=-0.43$ the gain curve $B_G$, and not $B_F$, is the most appropriate for describing the early-stage growth of the $2x-$mode.\\
This is confirmed by the numerical results depicted in Fig.~\ref{figNumericsMode2a}, where the amplitude amplification factor $\hat{A}_{2x}(L,\nu)$ obtained by numerical solution of Eq.(\ref{TotalEquation2}) is displayed when the fiber length $L=5$. In  Fig.~\ref{figNumericsMode2a}c we observe that, around $\nu=-0.43$, $\hat{A}_{2x}(L,\nu)\approx B_G(\nu)-0.07$, that is to say $\hat{A}_{2x}$ is a copy of $B_G$ lowered of $-0.07\equiv L^{-1}Log(\tilde{w}_1[2])$, as predicted by Eq.(\ref{SpectralEstimation}). \\
On the contrary, when $L>15.8$ the gain $g_2$ is dominant for the $2x-$mode at $\nu=-0.43$: the gain curve $B_F$ is therefore the most appropriate for describing the second-stage growth (i.e. for $\xi>15.8$) of the $2x-$mode. This is confirmed by Fig.~\ref{figNumericsMode2b} where the amplification factor $\hat{A}_{2x}(L,\nu)$ is shown when $L=16$; in this case, around $\nu=-0.43$, $\hat{A}_{2x}(L,\nu)\approx B_F(\nu)-0.21$, with $-0.21\equiv L^{-1}Log(\tilde{w}_2[2])$ as predicted by Eq.(\ref{SpectralEstimation}).\\
Differently from the case of the $2x-$mode, the gain $g_2$ is always dominant for the $4x-$mode in the band of $B_G$ centered at $\nu=-0.43$, because in that band $\tilde{w}_1[4]<\tilde{w}_2[4]$ (see Fig.~\ref{figEgv}c). The numerical solution of Eq.(\ref{TotalEquation2}) confirms indeed that the gain curve $B_F$ is the most suitable at describing the IM-MI growth of the $4x-$mode even at early-stage ( see Fig.~\ref{figNumericsMode4}).\\

The results displayed in Figs.~\ref{figNumericsMode2a}-~\ref{figNumericsMode4} show a good agreement between the numerical simulations and the analytical estimation Eq.(\ref{SpectralEstimation}), which represents therefore a simple and powerful tool for describing the IM-MI growth of the modes.\\
In Fig.~\ref{figNumericsMode2a} we see that the $2x-$mode is amplified in the bands of $all$ the gain curves $B_{A...J}$, which confirms what previously observed, namely that each mode of the fiber is amplified in the whole IM-MI band and its amplification depends on the eigenvalues and eigenvectors of  ${\bf M}$, which in turns depend on the system parameters.\\
We also point out that the term $L^{-1}Log(|\tilde{w}_{dom,nx}[n]|)$ in Eq.(\ref{SpectralEstimation}) explains the strong asymmetries in the amplification spectrum of the fiber modes which can be observed comparing panel (b) with (c) or (a) with (d) in Fig.~\ref{figNumericsMode2a}.
In fact the dominant gain at the signal and idler detuning $\pm\nu$ is the same, i.e. $g_{dom,nx}(L,\nu)=g_{dom,nx}(L,-\nu)$, but the corresponding eigenvector components are in general different, i.e. $\tilde{w}_{dom,nx}[n](L,\nu)\neq \tilde{w}_{dom,nx}[n](L,-\nu)$, which makes $\hat{A}_{nx}(L,\nu)\neq \hat{A}_{nx}(L,-\nu)$ according to Eq.(\ref{SpectralEstimation}).\\
To resume: the key concepts that should be retained from this section are essentially two. First, the modal amplification may exhibit a complex dynamics, which is due to the presence of different gains that are dominant at different fiber positions. Second, the amplification of each mode, well described by Eq.(\ref{SpectralEstimation}), can be controlled over the whole IM-MI band by means of the system parameters, which brings to the idea of wideband parametric amplification in multimode fibers.

\section{A physical insight into the IM-MI}\label{rif:inside}
\begin{figure}[htbp]
   \begin{center}
       \includegraphics[width=0.85\columnwidth]{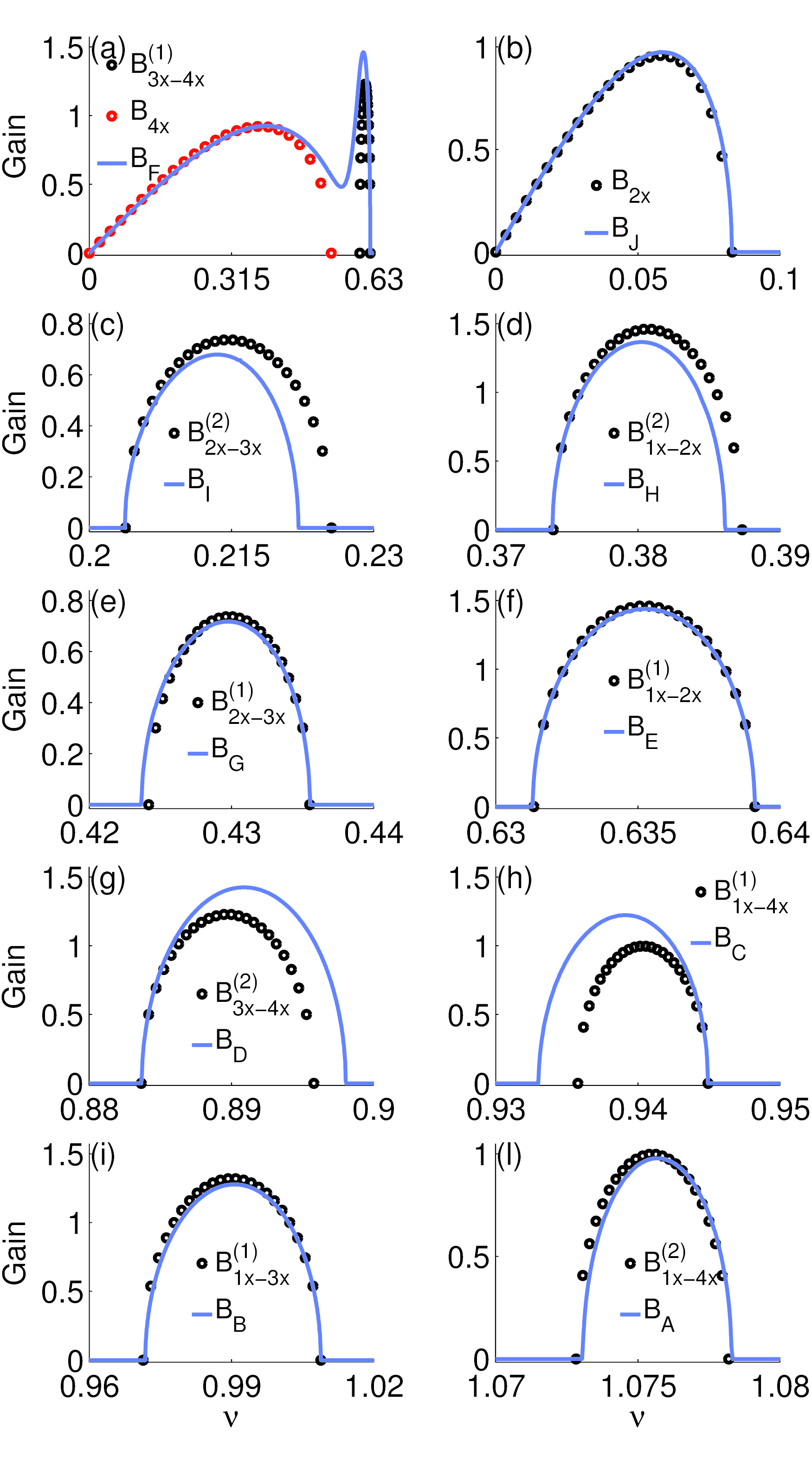}
   \end{center}
   \caption{Comparison between the gain curves $B_{A...J}$ computed from the full model Eq.(\ref{LinearSystem}) (blue solid lines) and the analytical estimations calculated from Eq.(\ref{SMISpectrum},~\ref{NaMbSpectrum})(black and red circles).}
   \label{ComparisonBands}
\end{figure}

The physical model resumed by Eq.(\ref{LinearSystem}) offers a semi-analytical solution that can precisely quantify the IM-MI growth in terms of eigenvalues and eigenvectors of the matrix ${\bf M}$.\\
A natural question that arises is how to control the IM-MI gain bands by means of the system parameters. From this point of view the model resumed by Eq.(\ref{LinearSystem}), although accurate, lacks simplicity as it makes necessary to compute at each detuning $\Omega$ the eigenvalues of ${\bf M}$.\\
For this reason in this section we aim at gaining a deeper physical insight by looking for some basic model that can easily explain the main features of the IM-MI process.\\
As outlined in the Introduction, a simple way of thinking the IM-MI in a multimode fiber is to consider separately the nonlinear interactions between all the possible couples of spatial and polarization modes. This approach, which we call bimodal-MI model, has been adopted in previous works concerning the IM-MI \cite{Stolen75,Mussot03,Tonello06}. In this section and in the next three sections we revise this basic model in order to find some useful analytical formulas for the IM-MI gain as well as to discuss the impact of the TOD and losses in the IM-MI dynamics.\\
If one takes into account for the interaction between the $na$-mode and the $mb$-mode, then Eq.(\ref{SidebandEquation}) should be rewritten by neglecting all the terms except for those involving the sidebands $s_{na}$,$s_{mb}$,$i_{na}$,$i_{mb}$:

\begin{flalign}
&\frac{\partial {s_{na}}}{\partial z} =i(\phi_{na}(z)+ b_S C_{nn}|p_{na}|^2)s_{na} +\nonumber\\
&i b_S C_{nn} p_{na}^2 i_{na}^*exp(-i\Delta\beta_{na}^{(p,s)}z-i\Delta\beta_{na}^{(p,i)}z)+&&\nonumber\\
& i b_{mn} C_{mn} p_{mb}^*p_{na} s_{mb} exp(i\Delta\beta_{mb}^{(p,s)}z-i\Delta\beta_{na}^{(p,s)}z)+\nonumber\\
& i b_{mn} C_{mn} p_{mb} p_{na} i_{mb}^* exp(-i\Delta\beta_{mb}^{(p,i)}z-i\Delta\beta_{na}^{(p,s)}z)
\label{SidebandBasicModel1}
\end{flalign}

where $b_{mn}=b_{||}$ if $a=b$ and $n\neq m$; $b_{mn}=b_{\bot}$ if $a\neq b$ and $n\neq m$; $b_{mn}=b_X$ if $a\neq b$ and $n=m$; $b_{mn}=0$ if $a=b$ and $n=m$. An equation similar to Eq.(\ref{SidebandBasicModel1}) could be written for $i_{na}$ by exchanging $s\leftrightarrow i$ , for $s_{mb}$ by exchanging $n\leftrightarrow m$ and $a\leftrightarrow b$, and for $i_{mb}$ by exchanging $s\leftrightarrow i$, $n\leftrightarrow m$ and $a\leftrightarrow b$.\\
The first term in the right-hand-side of Eq.(\ref{SidebandBasicModel1}) is here called phase-term as it is responsible for the phase modulation of  $s_{na}$. The second term is called self-MI-term as it accounts for the degenerate four-wave mixing (FWM) process where two $na-$pump photons are converted to an $na-$idler photon and a $na-$signal photon, leading to their amplification. Similarly, the last term is called cross-MI-term as it accounts for the FWM where a $na-$pump photon and a $mb-$pump photon are converted to an $na-$signal photon and a $mb-$idler photon. The third term is called not-MI-term as it does not account for sideband amplification.\\
We point out that if $a=b$ and $n=m$ we finally get a system of 2 equations for $s_{na}$ and $i_{na}$ that closely recall the system of equations describing the scalar MI (SMI) in a single-mode fiber \cite{AgrawalNLO01}; here we thus refer to this instance as $SMI_{na}$.\\
Otherwise, if $a\neq b$ and/or $n \neq m$ we get a system of 4 equations for $s_{na}$, $s_{mb}$, $i_{na}$ and $i_{mb}$ that recall the system of equations describing the vectorial MI (VMI) in single-mode fibers \cite{AgrawalNLO01}. Here we refer to this instance as $MI_{na-mb}$.\\
By generalizing the SMI in single-mode fibers \cite{AgrawalNLO01} to the most general case of the $SMI_{na}$ in a multimode fiber, we may easily infer that the gain curve $B_{na}$ related to the $SMI_{na}$ process reads as :

\begin{flalign}
& B_{na}=\frac{1}{2}|\beta_{2,na}|\Omega(\Omega_{c}^2-\Omega^2)^{1/2}\nonumber\\
& \Omega_{c}=2|p_{na}|\Big(\frac{b_S C_{nn}}{|\beta_{2,na}|}\Big)^{1/2}
\label{SMISpectrum}
\end{flalign}

which is valid under the anomalous dispersion condition $\beta_{2na}<0$ and where $\Omega_{c}$ is the upper cut-off frequency.\\
As regards the $MI_{na-mb}$ process, a rich and complex dynamics is found that strictly depends on the ratio $b_S/b_{mn}$ (see for example \cite{Seve96} for the VMI in single-mode Hi-Bi fibers).
The general way of handling the system of 4 equations for $s_{na}$, $s_{mb}$, $i_{na}$ and $i_{mb}$ consists in transforming it in an eigenvalue equation by means of a transformation similar to Eq.(\ref{Transformations}), and then to compute the corresponding dispersion relation. The problem is that, unless peculiar instances, the dispersion relation is a fourth-order polynomial equation for $\Omega$ requiring numerical computation to be solved.\\
On the other hand, different spatial modes typically exhibit a large group velocity mismatch (GVM) and the same occurs to the polarization modes in a Hi-Bi fiber. In this case the phase-matching conditions for the self-MI term, for the cross-MI-term and for the not-MI-term in Eq.(\ref{SidebandBasicModel1}) are reached at largely different detunings. For this reason, with the aim of studying the amplification induced by the cross-MI-term, we keep it along with the phase-term in Eq.(\ref{SidebandBasicModel1}), whereas we neglect the self-MI-term and the not-MI-term. Doing so, we can derive the following analytical estimate for the gain $B_{na-mb}$ related to the $MI_{na-mb}$ process (see Appendix II for details):


\begin{flalign}
&B_{na-mb}=\left(k_1^2 - k_2^2 (\Omega-\Omega_{PK})^2/\Omega_{PK,NL}^2 \right)^{1/2}\nonumber\\
&\Omega_{PK}= \Omega_{PK,L} + \Omega_{PK,NL}\nonumber\\
&\Omega_{PK,L}=\big|(2 D_{\beta 3})^{-1}\left(-\bar{\beta_{2}}\pm \big(\bar{\beta_{2}}^2-4 D_{\beta 3} D_v\big)^{1/2}\right)\big|  \nonumber\\
&\Omega_{PK,NL}=-2 k_2/(3\Omega_{PK,L}^2 D_{\beta_3} + 2\Omega_{PK,L}\bar{\beta_2} + D_v)\nonumber\\
&k_1=b_{mn} C_{mn}|p_{na}||p_{mb}|\nonumber\\
&k_2=(b_S/2) (C_{nn}|p_{na}|^2 + C_{mm}|p_{mb}|^2)
\label{NaMbSpectrum}
\end{flalign}


where $\Omega_{PK}$ stands for the peak-gain detuning that here is written as the sum between a linear contribution $\Omega_{PK,L}$ and a nonlinear contribution $\Omega_{PK,NL}$; $\bar{\beta_{2}}$ is the average GVD $(\beta_{2,na}+\beta_{2,mb})/2$; $D_{\beta 3}$ indicates the TOD difference $(\beta_{3,na}-\beta_{3,mb})/6$ and  $D_v$ is the $GVM \equiv v_{na}^{-1}-v_{mb}^{-1}$.\\
The $\pm$ operator in the third line of Eq.(\ref{NaMbSpectrum}) reveals that two distinct gain bands are generally found when $D_{\beta 3} \neq 0$, let us say $B_{na-mb}^{(1,2)}$. This issue is carefully addressed in the next section and clearly shows that the TOD, which does not affect MI in single-mode fibers, can play an important role in the IM-MI dynamics instead. \\
In Fig.~\ref{ComparisonBands} the gain curves $B_{na}$ and $B_{na-mb}$, computed according to the analytical estimations Eqs.(~\ref{SMISpectrum},~\ref{NaMbSpectrum}), are compared to the gain curves $B_{A...J}$ of the full model discussed in the previous section and displayed in Fig.~\ref{figGains}.\\
The two gain curves $B_{1x-2x}^{(1)}$ and $B_{1x-2x}^{(2)}$ clearly match with $B_E$ and $B_H$, respectively. Similarly $B_{1x-4x}^{(1)}$ and $B_{1x-4x}^{(2)}$ match respectively with $B_C$ and $B_A$; $B_{2x-3x}^{(1)} $ and $B_{2x-3x}^{(2)}$ match respectively with $B_G$ and $B_I$; $B_{3x-4x}^{(1)}$ and $B_{3x-4x}^{(2)}$ match respectively with the right-side of $B_F$ and with $B_D$.\\

The $SMI_{2x}$ and the $SMI_{4x}$ processes occur because of the anomalous dispersion conditions $\beta_{2,2x}<0$ and $\beta_{2,4x}<0$ (see Table I). Moreover, their corresponding bands $B_{2x}$ and $B_{4x}$ match with $B_J$ and the left-side of $B_F$, respectively.\\
We note the absence of the two gains related to the $MI_{2x-4x}$ process, which is due to the large TOD difference that makes $\bar{\beta_{2}}^2-4 D_{\beta 3}D_v<0$ and thus $\Omega_{PK,L}$ in Eq.(\ref{NaMbSpectrum}) not real-valued. As explained in next section, these bands may appear by introducing higher-order dispersion terms that we have not contemplated for the sake of simplicity.\\
The first of the two gain curves related to the $MI_{1x-3x}$ process, that is $B_{1x-3x}^{(1)}$, matches with $B_B$. The second is centered at $\nu=4.6$, out of the frequency window from $\nu=-1.1$ to $\nu=1.1$ displayed in Fig.~\ref{figGains}, and thus it is not represented.\\
It can be seen that the agreement between the approximated curves $B_{na}$, $B_{na-mb}$ and the gain curves $B_{A...J}$ is quite good at all detunings. This outcome confirms that we can primarily ascribe each one of the IM-MI gain bands to one of the possible $SMI_{na}$ and $MI_{na-mb}$ processes in the fiber. As a consequence, we can employ Eqs.(~\ref{SMISpectrum},~\ref{NaMbSpectrum}) to get an analytical estimation of the IM-MI gain.\\
Yet, we stress that some discrepancy is present between the approximated gain curves and $B_{A...J}$, which is due to having neglected the self-MI-term and/or the not-MI term in Eq.(\ref{SidebandBasicModel1}) as well as the interplay between all the fiber modes.\\
Indeed Eq.(\ref{SidebandBasicModel1}) describes the dynamics of 1 (if $n=m$ and $a=b$) or 2 (if $n\neq m$ or $a\neq b$ ) modes that we call phase-matched (PM) modes and that are amplified thanks to a phase-matched FWM , which is the self-MI-term or the cross-MI-term of Eq.(\ref{SidebandBasicModel1}). On the other hand, Eq.(\ref{SidebandBasicModel1}) totally disregards the other modes, that we call not-PM modes.\\
As already stressed, the not-PM modes also undergo amplification and their dynamics can be described only by means of the full model Eq.(\ref{LinearSystem}). Roughly speaking, the not-PM modes are amplified because they are nonlinearly coupled to the PM modes by means of a not-PM FWM.\\
Let us consider for example the gain band $B_C$, which is related to the $MI_{1x-4x}^{(1)}$ process and where the sidebands of the $1x-$mode and the $4x-$mode are amplified by means of a PM FWM process. The $2x-$mode does note take part to this PM FWM (it is a not-PM mode in the band of $B_C$) but even so it is nonlinearly coupled to the $1x-$mode and the $4x-$mode: indeed from Eq.(\ref{SidebandEquation}), written for the $2x-$signal, we find that $\partial_z s_{2x}$ depends on $p_{4x} p_{2x} i_{4x}^* exp(-i\Delta\beta_{4x}^{(p,i)}z-i\Delta\beta_{2x}^{(p,s)}z)$ and $p_{1x} p_{2x} i_{1x}^* exp(-i\Delta\beta_{1x}^{(p,i)}z-i\Delta\beta_{2x}^{(p,s)}z)$ . Although these last two FWM terms are not phase-matched in the band of $B_C$, still they lead to the amplification of $s_{2x}$ because the idler $i_{1x}$ and $i_{4x}$ grow exponentially.\\
We can therefore conclude that, although the basic model of Eq.(\ref{SidebandBasicModel1}) sheds an important light on the IM-MI dynamics and allows deducing some useful analytical approximation of the IM-MI gain curves, however it cannot describe the MI growth of the not-PM modes. At this purpose, the full-model Eq.(\ref{LinearSystem}) should be exploited, as it provides both the eigenvalues and the eigenvectors in order to correctly characterize the spatial evolution of each mode. Furthermore, it is only by means of the full-model that we can precisely calculate the IM-MI gain curves without any approximation.

\section{Impact of third-order dispersion: secondary MI bands}\label{rif:TOD impact}
\begin{figure}[htbp]
\hspace{0mm}%
       \includegraphics[width=1.05\columnwidth]{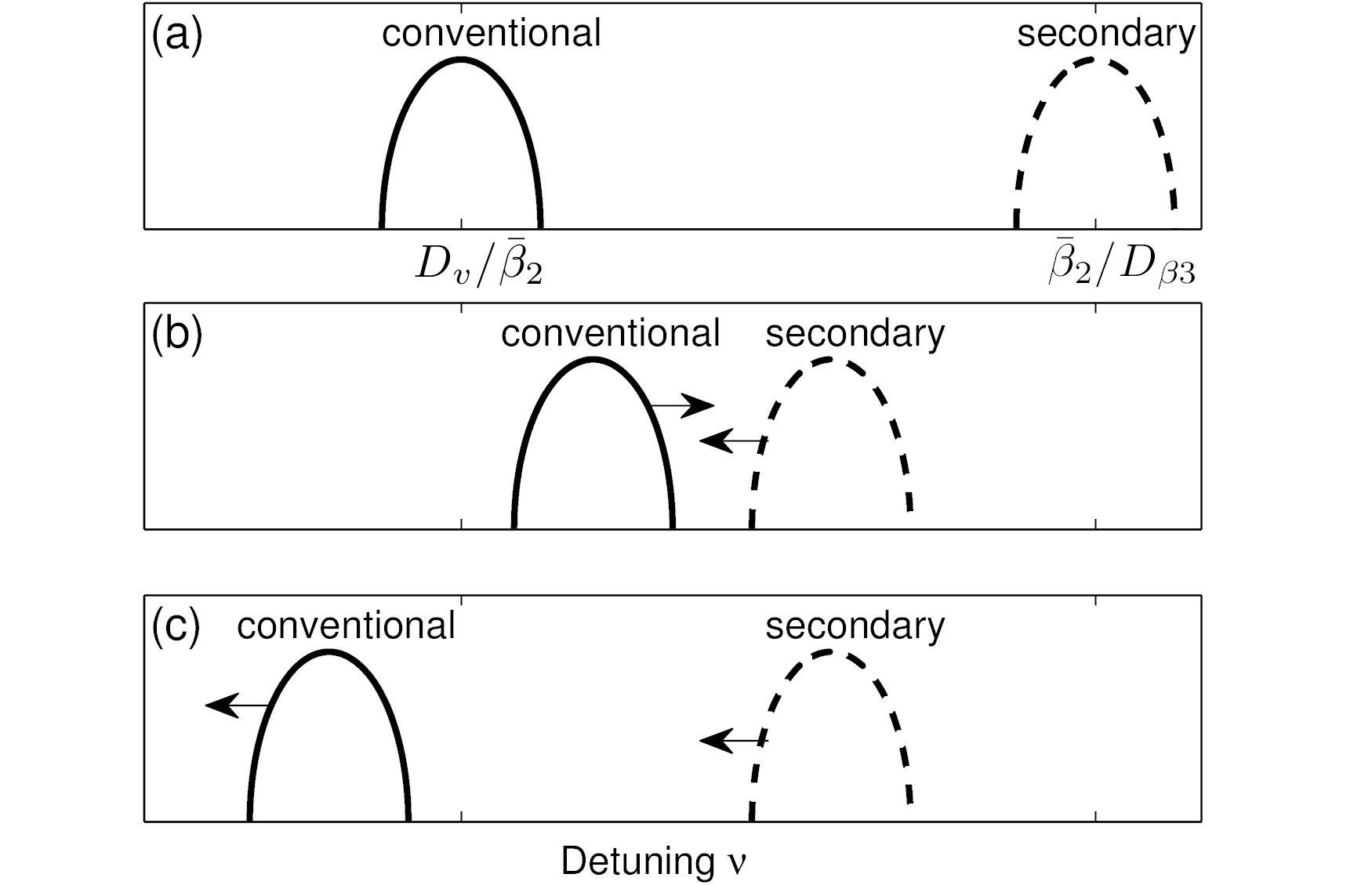}
 \caption{Schematic representation of TOD impact. Panel (a): conventional and secondary bands related to the $MI_{na-mb}$ process in the limit $|D_{\beta_3}|\rightarrow 0$. Panel(b): case of $sign(D_{\beta_3})=sign(D_v)$: for increasing values of $|D_{\beta_3}|$ the conventional band blue-shifts, whereas the secondary band red-shifts. Panel(c): case of $sign(D_{\beta_3})=-sign(D_v)$: for increasing values of $|D_{\beta_3}|$ both the bands red-shift. }
   \label{figAppendixTOD}
\end{figure}

Typically the large GVM between the $na-$mode and the $mb-$mode in the $M_{na-mb}$ process shifts the corresponding gain $B_{na-mb}$ towards high frequencies; consequently, the role of higher-order dispersion terms should be carefully analyzed. At the best of our knowledge, this has not been done in previous works concerning the bimodal-MI \cite{Stolen75,Mussot03,Tonello06}. However, a full treatment is complex and out of the scope of this paper, therefore in this section we just provide an outline of the influence of TOD on the IM-MI dynamics.\\
Equation (\ref{NaMbSpectrum}) reveals that two distinct bands are generally found when $|D_{\beta 3}|>0$ and it thus gives evidence of the remarkable impact of TOD on the IM-MI.\\
A similar behavior has been observed by Nithyanandan et al. in \cite{Porsezian12} in the case of two pumps co-propagating at different frequencies in a single-mode fiber. In that case a large TOD difference leading to the formation of 2 distinct bands was induced by a large frequency detuning between the two pumps, whereas in our case (single-pump) it originates from the different dispersion characteristics of different modes.\\
In order to better understand the role played by TOD, it is instructive to explore the limit $|D_{\beta 3}|\rightarrow 0$. From Eq.(\ref{NaMbSpectrum}) we see that in this limit the 2 bands are roughly centered around the peaks $\Omega_{PK}^{(1)}\approx |D_v/\bar{\beta_2}|$ and $\Omega_{PK}^{(2)}\approx |\bar{\beta_2}/D_{\beta_3}|$, respectively (we have neglected the nonlinear contribution $\Omega_{PK,NL}$) .\\
The first gain band, centered around $\Omega_{PK}^{(1)}$, is practically unaffected by the TOD. It is the conventional band discussed in previous works concerning the bimodal-MI.\\
The second band,centered around $\Omega_{PK}^{(2)}$, is strongly dependent on $D_{\beta_3}$ and represents a secondary MI spectrum in addition to the conventional one. It is typically located far away from the pump and for this reason higher-order dispersion terms besides the TOD could be not negligible.\\
According to Eq.(\ref{NaMbSpectrum}), the more $|D_{\beta_3}|$ increases the more the two bands move closer each other, but in a different way depending on the sign of $D_{\beta_3}$ and $D_v$ (see Fig.~\ref{figAppendixTOD}). If $sign(D_{\beta_3})=-sign(D_v)$ then both the bands move towards lower frequencies. Otherwise the conventional one moves towards higher frequencies and the secondary one towards lower frequencies, until they theoretically overlap once that $\beta_{2}^2=4 D_{\beta_3} D_v$. On the other hand, in the limit $\beta_{2}^2\rightarrow 4 D_{\beta 3} D_v$  the higher order terms besides the TOD play a not-negligible role, so that their influence on  the conventional and secondary band should be taken into account.

\section{Impact of the pump modal power distribution}\label{rif:pump impact}
As the higher-order dispersion terms, so the modal power distribution of the pump plays an important role in the IM-MI dynamics. This is evident considering that different power distributions produce different eigenvectors and eigenvalues of ${\bf M}$.\\
In this section we want to focus on the impact of the pump modal powers on the IM-MI gain bands.\\
Past works concerning the bimodal-MI \cite{Stolen75,Mussot03,Tonello06} have not put in evidence the nonlinear contribution $\Omega_{PK,NL}$ to the peak-gain detuning $\Omega_{PK}$ related to the $MI_{na-mb}$ process. Although generally $\Omega_{PK,NL}<<\Omega_{PK}$, however $\Omega_{PK,NL}$ has the same order of magnitude of the bandwidth of $B_{na-mb}$ and should therefore taken into account if we want to precisely locate this gain band. At this purpose we point out that, according to  Eq.(\ref{NaMbSpectrum}), $B_{na-mb}$ as function of $\Omega$ is a semi-ellipse centered in $\Omega_{PK}$ and whose bandwidth is $2\Omega_{PK,NL}(k_1/k_2)$.\\
It is also interesting to note that the peak-gain $k_1=b_{mn} C_{mn}|p_{na}||p_{mb}|$ of $B_{na-mb}$ is directly controlled by the two pump modal powers $|p_{na}|^2$ and $|p_{mb}|^2$ . Under the constraint $|p_{na}|^2+|p_{mb}|^2=constant$, $k_1$ is maximized when $|p_{na}|^2=|p_{mb}|^2$, namely, when the pump power is equally distributed over the $na-$mode and the $mb-$mode. \\
Contrary to the $MI_{na-mb}$, the $SMI_{na}$ process occurs without any GVM or TOD contribution to the phase-matching condition of the self-MI-term in Eq.(\ref{SidebandBasicModel1}). Indeed, phase-matching is achieved through direct compensation of second-order dispersion by self-focusing nonlinearity and as consequence both the position of the peak-gain detuning and the value of the peak-gain are strongly dependent on the pump power $|p_{na}|^2$.

\section{Impact of modal propagation losses}\label{rif:losses impact}
\begin{figure}[htbp]
\hspace{0mm}%
       \includegraphics[width=0.9\columnwidth]{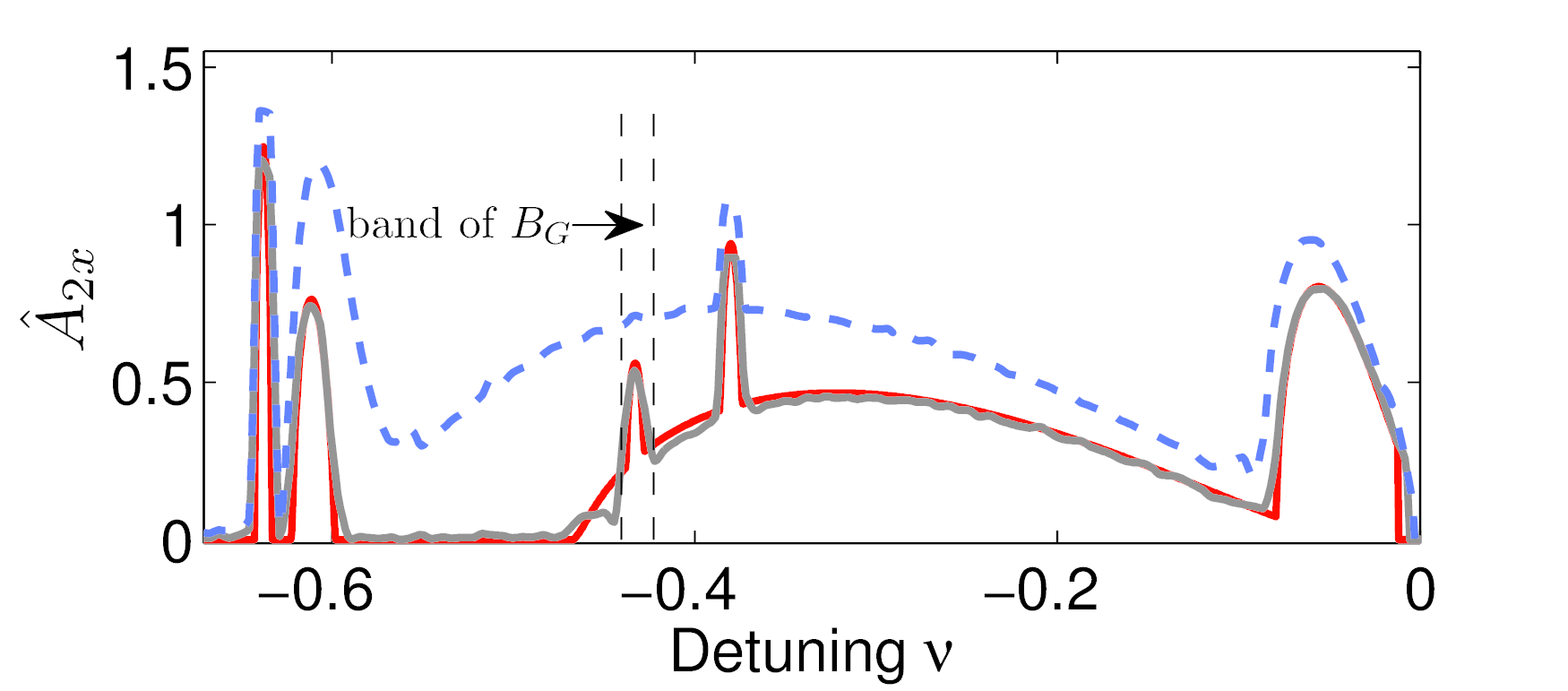}
 \caption{Amplitude amplification factor for the $2x-$mode when $L=16$ and in the band from $\nu=-0.65$ to $\nu=0$. In gray: $\hat{A}_{2x}$ obtained by numerical solution of Eq.(\ref{TotalEquation2}) by split-step method when propagation losses are included ($\alpha_{1x}=0.005$; $\alpha_{2x}=0.01$ ; $\alpha_{3x}=0.015$; $\alpha_{4x}=0.02$). In red: the estimation $\hat{A}_{2x,est}$ obtained by Eq.(\ref{SpectralEstimation}) using eigenvectors and eigenvalues of ${\bf M_{av}}$. In dashed blue: $\hat{A}_{2x}$ obtained by numerical solution of Eq.(\ref{TotalEquation2}) but with zero losses.}
   \label{figAppendix3}
\end{figure}

Typically higher-order spatial modes suffer larger attenuation than lower order modes, which makes interesting to study the influence of losses in the context of IM-MI.\\
Past works concerning the MI in single mode lossy fibers have demonstrated that the MI dynamics could be significantly influenced by fiber losses(\cite{Anderson84,Karlsson95,Dinda07}): indeed an optimum propagation length exists for which the MI growth is maximized and beyond which it is gradually annihilated by losses, and the MI spectrum experiences a continual frequency shifts towards the pump.\\
In this section for simplicity of notation we assume that losses are not dependent on $\Omega$; however $\Omega-$dependent losses could be easily treated.\\
Modal losses are included by adding the terms $i\alpha_{na}$ and $-i\alpha_{na}$ to the elements ${\bf M}_{sa,sa}[n,n]$ and ${\bf M}_{ia,ia}[n,n]$ of the matrix ${\bf M}$ defined in Section ~\ref{rif:general}, respectively ( $1\leq n\leq N$,$a=\{x,y\}$ ).  In addition, losses cause an exponential decay of the pump amplitudes, therefore the terms $|p_{na}|$ of the matrix ${\bf M}$ should be replaced by $|p_{na}(z=0)|exp(-\alpha_{na}z)$. In this way the coefficients of the matrix ${\bf M}$ are no longer constant, which prevents the existence of a simple solution to the eigenvalue problem $\partial_z {\bf v}=i{\bf M}{\bf v}$ of Eq.(\ref{LinearSystem}).\\
On the other hand we are interested to the regime in which the z-dependent coefficients $|p_{na}(0)|exp(-\alpha_{na}z)$ are slowly decaying if compared to the fast MI-growth of the sidebands, otherwise a fiber-amplifier would be low-efficient. In this regime the solution ${\bf v}$ could be well approximated by a Magnus series expansion truncated at the first order (\cite{Magnus54}), that is ${\bf v}(L)=exp(i {\bf M_{av}}L){\bf v}(0)$, with ${\bf M_{av}}=L^{-1}\int_0^L {\bf M}(z)dz$.\\
This means that when losses are taken into account then the vector ${\bf v}(L)$ is still obtained by Eq.(\ref{EvolutionV}) provided that we compute the eigenvalues $\lambda_k$ and the eigenvectors $w_k$ of the matrix ${\bf M_{av}}$, which is constructed from the matrix ${\bf M}$ defined in Section ~\ref{rif:general} by replacing the terms $|p_{na}|$ with their spatial average $L^{-1}\int_0^L|p_{na}(0)|exp(-\alpha_{na}z)dz$ and by adding the loss coefficients $i\alpha_{na}$ and $-i\alpha_{na}$ as previously explained.\\
{\bf Note that large losses may prevent the existence of MI phenomena: in a single-mode fiber, this corresponds to the case in which propagation losses are larger than the MI gain. We should therefore wonder if a given input pump power distribution allows or not for IM-MI. The response comes from the eigenvalues of the modified matrix ${\bf M_{av}}$ evaluated at the fiber entry (i.e. fixing $L=0$): if at least one eigenvalue with negative imaginary part exist, then IM-MI will occur}.\\
In order to assess the validity of our approach, we repeat the simulation described in Section ~\ref{rif:simulations} but including the following amplitude losses (per nonlinear length $L_{NL,1}$): $\alpha_{1x}=0.005$; $\alpha_{2x}=0.01$ ; $\alpha_{3x}=0.015$; $\alpha_{4x}=0.02$. In the case under analysis such losses are much larger than typical losses in silica fibers at telecommunication wavelengths($\alpha_{1x}=0.005$ corresponds to a real power loss of about 0.43 dB/m),nonetheless they are small compared to the IM-MI gain; moreover, here our purpose is to show the robustness of our approach even in presence of a relatively fast pump decaying.\\
In Fig.~\ref{figAppendix3} we show the good agreement between the amplitude amplification factor $\hat{A}_{2x}$, obtained by numerical simulations, and its estimate $\hat{A}_{2x,est}$, obtained by Eq.(\ref{SpectralEstimation}) where the dominant gain and the corresponding eigenvector are those associated to the matrix ${\bf M_{av}}$.\\
In this example losses have a deep impact over the MI growth of the $2x-$mode. We have seen in Section ~\ref{rif:simulations} that in absence of losses the gain $g_2=0.9$ is larger than $g_1=0.71$, which makes $g_2$ to be dominant, for the $2x-$mode, starting from $L=15.8$. For this reason, in $L=16$ and around $\nu=-0.43$, $\hat{A}_{2x}\approx B_F + L^{-1}Log(|\tilde{w}_{2}[2]|)$ (see dashed blue line in the band of $B_G$ in Fig.~\ref{figAppendix3}). On the other hand when losses are included, we find that for $L=16$ the new gains $g_2$ and $g_1$, calculated from ${\bf M_{av}}$, are respectively $g_2=0.45$ and $g_1=0.56$, which makes $g_1$ to be dominant instead of $g_2$ and therefore $\hat{A}_{2x}\approx B_G + L^{-1}Log(|\tilde{w}_{1}[2]|)$ (Fig.~\ref{figAppendix3}, gray line).\\
Following a treatment similar to that exposed in Appendix II the estimate of $B_{na-mb}$ given by Eq.(\ref{NaMbSpectrum}) could be recalculated by taking into account for losses. The matrix ${\bf M'}$ of Appendix II should be rewritten by adding the loss coefficients and by replacing the pump amplitudes with their corresponding spatial averages.\\
After some algebra we find that the peak-gain $B_{na-mb,PK}^{(ls)}$ may be approximated as:

\begin{flalign}
B_{na-mb,PK}^{(ls)}=B_{na-mb,PK}\Big(1-exp(-2\bar{\alpha} L)\Big)/(2\bar{\alpha} L) - \bar{\alpha}
\label{app3_1}
\end{flalign}

where the superscript $(ls)$ stands for losses, whereas $B_{na-mb,PK}$ is the peak-gain in absence of losses and $\bar{\alpha}$ indicates the average $(\alpha_{na}+\alpha_{mb})/2$. Interestingly enough, the drop of the peak-gain is thus related to the average between the loss coefficient of the $na-$mode and the loss coefficient of the $mb-$mode.\\
A similar equation holds true for the peak-gain $B_{na,PK}^{(ls)}$ of the the $SMI_{na}$ process, after replacement of $B_{na-mb,PK}$ with $B_{na,PK}$ and of $\bar{\alpha}$ with $\alpha_{na}$ in Eq.~\ref{app3_1}. \\
Furthermore, the peak-gain detuning related to the $SMI_{na}$ undergoes a red-shift, which is completely similar to the MI frequency drift already predicted in single-mode fibers and which can be estimated as follows:

\begin{flalign}
\Omega_{PK}^{(ls)}=\Omega_{PK}\Big(1-exp(-\alpha_{na} L)\Big)/(\alpha_{na} L)
\label{app3_2}
\end{flalign}

On the contrary, in the $MI_{na-mb}$ processes only the nonlinear contribution $\Omega_{PK,NL}$ is shifted by losses, but not the linear contribution $\Omega_{PK,L}$ which is significantly larger than $\Omega_{PK,NL}$. As a consequence,the global peak shift induced by losses is typically negligible.\\

\section{Limits of validity}\label{rif:limits}
In this section we discuss the limits of validity of the main results deduced in this paper.\\
We remind that in Eq.(\ref{TotalEquation}) several nonlinear terms have been neglected because considered fast-oscillating. As an example, let us take the nonlinear term $c_{nnmn}P_{nx}^2P_{mx}^*e_{1,nnmn}$, with $n\neq m$ and $c_{nnmn}\neq 0$, whose corresponding wavevector mismatch is $\Delta\beta_{1,nnmn}\equiv \beta_{nx}(\omega_p)-\beta_{mx}(\omega_p)$. Typically the propagation constants $\beta_{nx}(\omega)$ and $\beta_{mx}(\omega)$ are largely different whatever the frequency $\omega$ is, which justifies the statement $|\Delta\beta_{1,nnmn}L_{NL}|>>0$. On the other hand, for a proper fiber design these 2 propagation constants could become equal at a peculiar frequency, let us say $\bar{\omega}$. If this is the case and the pump frequency $\omega_p$ coincides with $\bar{\omega}$, then this nonlinear term must be taken into account in Eq.(\ref{TotalEquation}). What said is generally true for anyone of the terms that we have neglected.\\
Note also that, although the neglected terms are typically truly fast oscillating, however they could lead to some MI process that we have not accounted in our analysis. As an example, let us consider once again the term $c_{nnmn}P_{nx}^2P_{mx}^*e_{1,nnmn}$ previously discussed, assuming $\omega_p \neq \bar{\omega}$ so that it is effectively fast-oscillating. Its decomposition by means of Eq.(\ref{Decomposition}) would lead to the presence of the term $c_{nnmn}p_{nx}^2i_{mx}^*exp(i2\beta_{p,nx}-i\beta_{i,mx}-i\beta_{s,nx})$ in Eq.(\ref{SidebandEquation}), which is associated to a FWM process where energy is transferred from the $nx-$pump to the $nx-$signal and the $mx-$idler. Once again, these considerations are generally true for anyone of the terms that we have dismissed in Eq.(\ref{TotalEquation}); consequently, a group of MI bands exist that we have ignored in our analysis. However, these MI bands are typically located far away from the gain bands predicted by our theory; moreover the corresponding FWM processes are very sensitive to small fiber imperfections so that their typical coherence length is less than 1 meter (\cite{Stolen75}), which practically prevent an efficient MI growth in fibers longer than a few meters.\\
For this reason we can safely assume that these neglected MI-bands do not affect the analysis developed in this paper. On the other hand, for very high-input powers and/or a large number of propagating modes in short fibers, some of the IM-MI gain bands predicted by our analysis would be located far away from the pump and could thus overlap with the neglected MI-bands, which may invalidate our results in the portion of the spectrum where overlap occurs.\\
Another important issue that we have not considered in our analysis is the effect of polarization mode dispersion (PMD) as well as of spatial mode dispersion (SMD) related to groups of quasi-degenerate modes, which could greatly impair the IM-MI growth in telecom fibers \cite{Guasoni12}. Indeed, in the case of telecom fibers, Eq.(\ref{SidebandEquation}) is strictly valid in the regime of zero-PMD and zero-SMD only.\\
A proper analysis taking into account for the PMD and SMD could be done following a treatment similar to that proposed in \cite{Guasoni12} in the case of single-mode fibers. The final result would be the presence of exponentially-decaying nonlinear coefficients in Eq.(\ref{SidebandEquation}), which is completely analogue to the presence of exponentially decaying pump amplitudes when propagation losses are taken into account. We could therefore employ the Magnus expansion proposed in Section ~\ref{rif:losses impact} to deal with both losses and PMD/SMD. On the other hand, the derivation of the exponentially-decaying nonlinear coefficients in presence of PMD and SMD is complex and out of the scope of this paper.

\section{Conclusions and perspectives}\label{rif:conclusions}
In this paper we have presented a detailed theoretical and numerical analysis of the IM-MI in multimode fibers, which brings to the idea of wideband multimode parametric amplification. We can summarize the main results as follows:\\
\\
{-\it Eigenvalue equation for multimode fibers-} One of the main outcomes of this work consists in the linearization of the coupled nonlinear  Schr\"{o}dinger equations describing the propagation in a multimode fiber and in their following transformation in an eigenvalue problem. In this way the modal amplification can be described by means of the eigenvectors and eigenvalues of a matrix ${\bf M}$ that contains the information about the dispersion characteristics of the modes and the modal power distribution of the pump. This result generalizes previous studies of MI with 2 spatial or polarization modes to the most generale case of $N>2$ interacting modes. The computation of the eigenvectors and eigenvalues of ${\bf M}$ as function of the pump-sideband detuning $\Omega$ permits to completely characterize the IM-MI. In particular, the eigenvalues with negative imaginary part, called as usual gains, lead to sideband amplification.\\
\\
{-\it Dominant gain and wideband amplification-}  For a given detuning several gains could be found, each one playing the role of dominant gain at different fiber positions and for different fiber modes. The amplification of each mode is thus controlled by its dominant gain as well as the corresponding eigenvector, which in turn depend on the system parameters. This issue is well highlighted by Eq.(\ref{SpectralEstimation}), which offers an analytical estimate of the modal amplification. Most importantly, as highlighted in  Section ~\ref{rif:general}, each mode undergoes amplification in the whole IM-MI band.\\
 \\
{-\it Physical interpretation of the multimode amplification-} Although the eigenvalues and eigenvectors of ${\bf M}$ can completely characterize the multimode amplification, however they do not provide a simple and intuitive picture of the physics behind the IM-MI. At this purpose we considered separately the nonlinear interactions between all the possible couples of modes, which allows decomposing the complex IM-MI dynamics in a set of bimodal-MI processes. The bimodal nonlinear interaction between the $na-$mode and the $mb-$mode gives rise to a MI process, called $MI_{na-mb}$, completely analogous to the vectorial MI in single-mode fibers. Similarly the self nonlinear interaction of the $na-$mode gives rise to the $SMI_{na}$ process, which is analogous to the scalar MI in single-mode fibers. These analogies permit to find out the analytical estimates Eqs.(~\ref{SMISpectrum},~\ref{NaMbSpectrum}) approximating the IM-MI gain and to interpret the wide IM-MI band as the union of the bands related to each $MI_{na-mb}$ and $SMI_{na}$ process. As a result, the useful bandwidth is in principle much larger than in the case of single-mode fibers.\\
 \\
{-\it Phase-matched and not-phase-matched modes-} In the simple bimodal model describing the $MI_{na-mb}$ and the $SMI_{na}$ processes the $na-$mode and the $mb-$mode play the role of phase-matched modes, namely, they are amplified by means of a phase-matched FWM. Nevertheless, this simple model cannot give any information about the other modes, which we call not-phase matched and which undergo amplification because they are nonlinearly coupled to the phase-matched modes. Therefore, although it sheds an important light on the IM-MI dynamics, the bimodal model does not offer a complete characterization of the IM-MI amplification, which is provided by the eigenvectors and the eigenvalues of ${\bf M}$ instead.\\
 \\
{-\it Influence of TOD and losses-} We have concluded our work by analyzing the impact of TOD and of losses on the IM-MI dynamics. We have put in evidence that the large GVM between different spatial modes may shift the IM-MI gain bands towards high frequencies, so that higher-order dispersion terms should be contemplated in order to correctly describe the IM-MI. Differently from single-mode fibers, the TOD difference between different modes plays an important role and leads to the formation of secondary MI-bands in addition to the conventional ones.\\
Contrary to the GVM, propagation losses may shift the gain bands towards low frequencies. Their impact (gain drop and frequency shift) can be evaluated by properly modifying the matrix ${\bf M}$ and resorting to a Magnus expansion which accounts for the spatial decaying of the pump amplitudes.\\
\\
{-\it Future perspectives-} The outcomes exposed in this paper pave the way towards the implementation of wideband multimode parametric amplifiers where the amplification of each mode could be selectively controlled by means of the system parameters. For this reason a natural evolution of this work is the study of the optimization of the pump modal distribution and of the fiber parameters in order to maximize the amplification of one or more modes in a desired band. {\bf A further important issue to address will concern the system scalability}. In addition, although limited to the case of single-core step-index fibers, the theory here developed could be conveniently modified so as to describe IM-MI phenomena in multicore fibers or photonic crystal fibers, which provide a unique opportunity for tailoring the modal dispersion characteristics and the nonlinear coupling coefficients. {\bf Finally, this work may find a useful application in the description of the complex spatio-temportal soliton dynamics in multimode fibers, which recently has been widely discussed \cite{Wright15,Buch15,Picozzi15} and represents a truly hot-topic in optical-physics.}

\section*{Acknowledgments}
I thank A.Picozzi and G.Millot for fruitful discussions. This work was supported by the European Research
Council under Grant 306633, ERC PETAL.

\renewcommand{\theequation}{A1-\arabic{equation}}
\setcounter{equation}{0}  
\renewcommand{\thefigure}{A1-\arabic{figure}}
\setcounter{figure}{0}
\section*{Appendix 1}
\begin{figure}[htbp]
   \begin{center}
       \includegraphics[width=0.9\columnwidth]{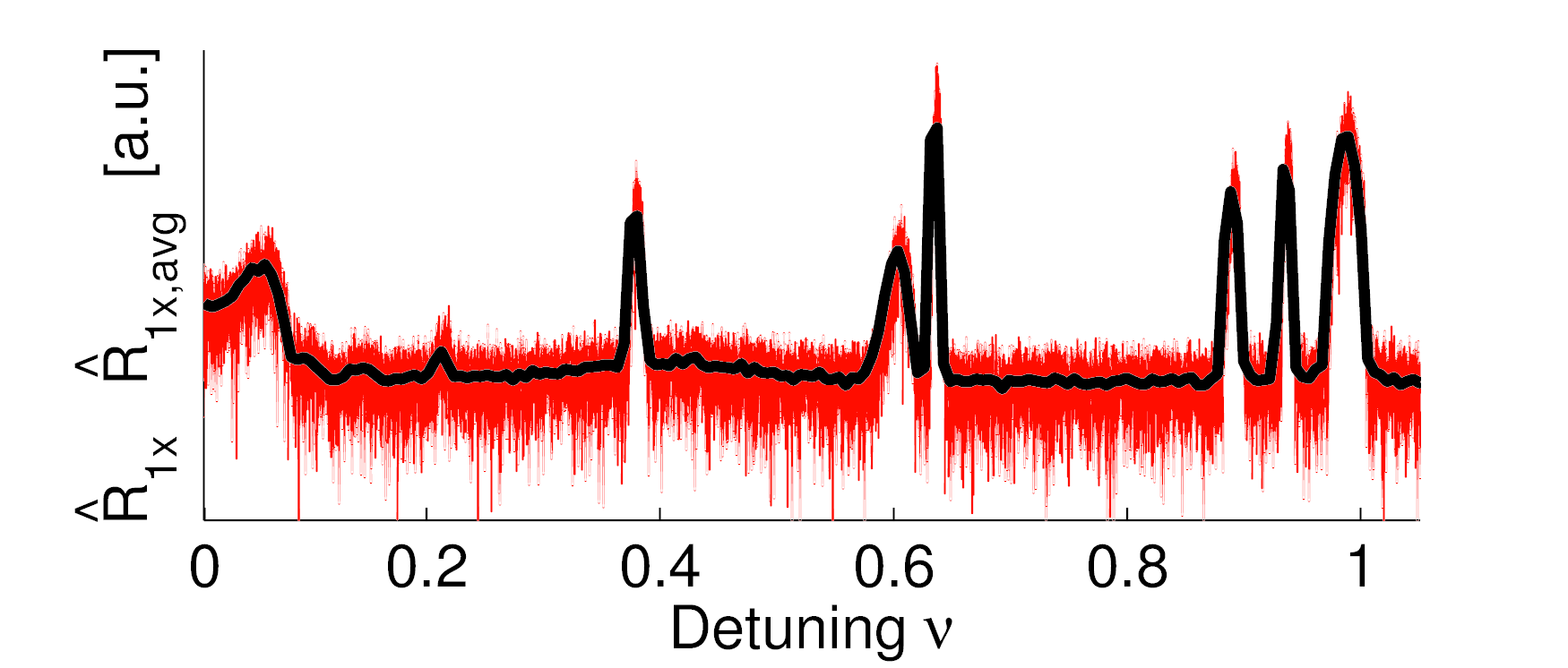}
   \end{center}
   \caption{Comparison between $\hat{R}_{1x}$ (red) and $\hat{R}_{1x,avg}$ (black).}
   \label{figAppendix}
\end{figure}

In this Appendix we derive Eq.(\ref{SpectralEstimation}), which represents an analytical estimate for the amplification factor related to a weak input background noise amplified by the IM-MI process.\\
Each input modal field is the sum between a pump component $p_{na}(0,t)$ and a noise $r_{na}(0,t)$. For the sake of clarity in this Appendix we refer to the $1x-$mode but what follows can be easily generalized to anyone of the propagating modes.\\
The power spectrum $\hat{R}_{1x}(z,\nu)\equiv |\mathcal{F}\{r_{1x}(z,t)\}|^2$ ($\mathcal{F}$ indicates the Fourier transform) typically exhibits strong and fast fluctuations due to the random nature of $r_{1x}$ (Fig.~\ref{figAppendix}).
{\bf In order to correctly estimate the noise spectrum, and thus to reduce the spectral fluctuations, a statistical average should be performed over several realizations of input noise. Clearly, this approach is extremely time consuming. An alternative option, which turns out to be equivalent to the aforementioned approach, consists in averaging the power spectrum in a narrow band of width $b$ where several fluctuations are included}, so that the averaged power spectrum reads as  $\hat{R}_{1x,avg}(z,\nu)=b^{-1}\int_{\nu-b/2}^ {\nu+b/2} \hat{R}_{1x}(z,\psi)\partial\psi $ (Fig.~\ref{figAppendix}).\\

{\bf This averaging process is practically equivalent to applying a moving-average filter to  $\hat{R}_{1x}(z,\nu)$: it allows suppressing} the fast fluctuations in the spectrum and recovering the real noise power level around a certain frequency.\\
In order to find an estimate of the amplification factor $\hat{A}_{1x}$, we start from Eq.(\ref{EvolutionV}) written for the $1x-$mode, namely:

\begin{flalign}
v[1](z,\nu) = \sum_{k=1}^{4N} c_k w_k[1] exp(i\lambda_k z)
\label{v1}
\end{flalign}

where $v[1](z,\nu)$ indicates the amplitude of the $1x-$mode at the position $z$ and at the detuning frequency $\nu$, that is $|v[1](z,\nu)|^2 \equiv \hat{R}_{1x}(z,\nu)$, and the wavevectors are normalized so that $|{\bf w}_{k}\bullet{\bf w}_{k}| =1$. For notational simplicity, in this Appendix we omit to indicate the dependence of $c_k$, ${\bf w_k}$ and $\lambda_k$ on $\nu$, although they are actually $\nu-$dependent; indeed ${\bf w_k}(\nu)$ and $\lambda_k(\nu)$ ($1\leq k \leq 4N$) are respectively the eigenvectors and the eigenvalues of the matrix ${\bf M}$ computed at the frequency $\nu$, whereas $c_k(\nu)=( {\bf v}(0,\nu)\bullet{\bf D}{\bf w_k} )/ ( {\bf w_k}\bullet{\bf D}{\bf w_k} )$. We also omit to indicate the interval of integration, so we indicate $\int_{\nu-b/2}^ {\nu+b/2}$  simply with $\int$.\\
The elements $v[n](0,\nu)$ and $v[n+N](0,\nu)$ of ${\bf v}(0,\nu)$ ($1\leq n \leq N$) represent the input noise amplitude for the $nx-$mode and the $ny-$mode at the Stokes frequency $\nu$, respectively, whereas $v[n+2N](0,\nu)$ and $v[n+3N](0,\nu)$ are the noise amplitudes for the $nx-$mode and $ny-$mode at the anti-Stokes frequency $-\nu$; therefore $|v[n](0,\nu)|^2 \equiv \hat{R}_{nx}(0,\nu)$, $|v[n+N](0,\nu)|^2 \equiv \hat{R}_{ny}(0,\nu)$, $|v[n+2N](0,\nu)|^2 \equiv \hat{R}_{nx}(0,-\nu)$, $|v[n+3N](0,\nu)|^2 \equiv \hat{R}_{ny}(0,-\nu)$.
Furthermore, the integrals $b^{-1}\int |v[n](0,\psi)|^2\partial\psi\equiv$$\hat{R}_{nx,avg}(0,\nu)$ and $b^{-1}\int |v[n+N](0,\psi)|^2\partial\psi\equiv$$\hat{R}_{ny,avg}(0,\nu)$ indicate respectively the input noise power related to the $nx-$mode and to the $ny-$mode in the Stokes band $[\nu-b/2,\nu+b/2]$; similarly $b^{-1}\int |v[n+2N](0,\psi)|^2\partial\psi\equiv$$\hat{R}_{nx,avg}(0,-\nu)$ and $b^{-1}\int |v[n+3N](0,\psi)|^2\partial\psi\equiv$$\hat{R}_{ny,avg}(0,-\nu)$ indicate the input noise powers in the anti-Stokes band $[-\nu-b/2,-\nu+b/2]$.\\
Since the input background noises are white and independent of each other, we can safely assume that the averaged spectra $\hat{R}_{nx,avg}(0,\nu)$ and $\hat{R}_{ny,avg}(0,\nu)$ are independent of the polarization, of $n$ and $\nu$, so that $\hat{R}_{na,avg}(0,\pm\nu)$$\approx$$\hat{R}_{1x,avg}(0,\nu)$ ($a=\{x,y\}, 1\leq n \leq N$) and therefore $\int |v[k](0,\psi)|^2\partial\psi$$\approx$$\int |v[1](0,\psi)|^2\partial\psi$ $\equiv$ $b\hat{R}_{1x,avg}(0,\nu)$   ($\forall k:$ $ 1\leq k \leq 4N$).\\
We may approximate Eq.(\ref{v1}) by taking into account only the dominant gain $g_{dom,1}$ for $v[1]$ at the position $z$ and the associated eigenvector ${\bf w}_{dom,1}$, that is $v[1](z,\nu) \approx c_{dom,1} w_{dom,1}[1] exp(g_{dom,1}z)$.
In this way we could rewrite $\hat{R}_{1x,avg}(z,\nu)$ $\equiv$ $b^{-1}\int |v[1](z,\psi)|^2\partial\psi$=$b^{-1}p_1 \int |c_{dom,1}|^2\partial\psi$, where $p_1=|w_{dom,1}[1]|^2 exp(2 g_{dom,1}z)$ and we have assumed $ w_{dom,1}[1]$ and $ g_{dom,1}$ to be constant in the small band $[\nu-b/2,\nu+b/2]$ so that they can be taken outside the integral.\\
When calculating $\int |c_{dom,1}|^2\partial\psi$ we should take into account that couples of different input modal noises are independent each other, therefore $\int v[n](0,\nu)\cdot v[m]^*(0,\nu)\partial\psi\approx 0$ if $n\neq m$ and then $\int |c_{dom,1}|^2\partial\psi \approx$ $p_2\sum_{k=1}^{4N} |w_{dom,1}[k]|^2 \int |v[k](0,\psi)|^2\partial\psi$, with $p_2=|{\bf w}_{dom,1}\bullet{\bf D}{\bf w}_{dom,1}|^{-2}$.\\
Being $\sum_{k=1}^{4N} |w_{dom,1}[k]|^2$$\equiv$ $|{\bf w}_{dom,1}\bullet{\bf w}_{dom,1}| =1$ and $\int |v[k](0,\psi)|^2\partial\psi$ $\approx$ $b\hat{R}_{1x,avg}(0,\nu)$, we conclude that $\int |c_{dom,1}|^2\partial\nu$=$p_2 b\hat{R}_{1x,avg}(0,\nu)$, and then $\hat{R}_{1x,avg}(z,\nu)=p_1p_2 \hat{R}_{1x,avg}(0,\nu)$, from which Eq.(\ref{SpectralEstimation}) is easily derived.

\renewcommand{\theequation}{A2-\arabic{equation}}
\setcounter{equation}{0}  
\renewcommand{\thefigure}{A2-\arabic{figure}}
\setcounter{figure}{0}
\section*{Appendix 2}
In this Appendix we derive Eq.(\ref{NaMbSpectrum}), which represents an estimate for the gain $B_{na-mb}$ related to the $MI_{na-mb}$  process.\\
As suggested in Section ~\ref{rif:inside}, in Eq.(\ref{SidebandBasicModel1}) we may just retain the phase-term and the cross-MI-term. As a result, the system of 4 CNLSE describing the dynamics of $s_{na},i_{na},s_{mb},i_{mb}$ is splitted in two distinct subsystems: the first accounting for the interaction between $s_{na}$ and $i_{mb}$; the second for the interaction between $s_{mb}$ and $i_{na}$. We indicate with $MI_{sna-imb}$ and $MI_{ina-smb}$ the corresponding MI processes.\\
As regards the $MI_{sna-imb}$: making use of Eqs.(~\ref{PumpSolution},~\ref{Transformations}) we finally get the eigenvalue problem $\partial_z {\bf v'}=i{\bf M'}{\bf v'}$,  being ${\bf v'}=[\bar{s}_{na}\,\, \bar{i}_{mb}^*]^T$ and ${\bf M'}$ the 2x2 matrix whose elements are respectively ${\bf M'}[1,1]={{\bf M_{sa,sa}}[n,n]}$, ${\bf M'}[2,2]=-{{\bf M_{ib,ib}}[m,m]}$ and ${\bf M'}[1,2]=-{\bf M'}[2,1]={{\bf M_{sa,ib}}[n,m]}$ (matrix ${\bf M}$ is defined in Section ~\ref{rif:general}) .\\
The gain $B_{na-mb}$ is given by the eigenvalues $\lambda$ of ${\bf M'}$ with negative imaginary part, namely:

\begin{flalign}
B_{na-mb}=\Big({\bf M'}[1,2]^2 -4^{-1}({\bf M'}[1,1]-{\bf M'}[2,2])^2 \Big)^{1/2}
\label{app2_1}
\end{flalign}

According to Eq.(\ref{app2_1}) the peak-gain ${\bf M'}[1,2]\equiv b_{mn} C_{mn}|p_{na}||p_{mb}|$ is reached at the peak-gain detuning $\Omega_{PK}$ for which the phase-matching condition ${\bf M'}[1,1]-{\bf M'}[2,2]=0$ occurs. The term ${\bf M'}[1,1]-{\bf M'}[2,2]\equiv L+NL$ is the sum of a linear part $L$ and a nonlinear part $NL$ that read as:

\begin{flalign}
&NL=b_S(C_{nn}|p_{na}|^2+C_{mm}|p_{mb}|^2)\nonumber\\
&L(\Omega)=\Delta\beta_{na}^{(p,s)} + \Delta\beta_{mb}^{(p,i)}\equiv
  D_{\beta_3}\Omega^3 + \bar{\beta_2}\Omega^2 + D_v\Omega
\label{app2_2}
\end{flalign}

where $\Delta\beta_{na}^{(p,s)}$ and $\Delta\beta_{mb}^{(p,i)}$ are given by Eq.(\ref{TaylorExpansion}), $D_{\beta 3} =(\beta_{3,na}-\beta_{3,mb})/6$, $\bar{\beta_{2}}=(\beta_{2,na}+\beta_{2,mb})/2$ and  $D_v=v_{na}^{-1}-v_{mb}^{-1}$. Note that according to this notation $\Omega$ is positive-valued. In this Appendix we assume that the coupling coefficients $C_{nn}$ and $C_{mm}$ are independent of $\Omega$ in the band of $B_{na-mb}$, therefore $NL$ is independent of $\Omega$, too. \\
The $na-$mode and the $mb-$mode that are involved in the $MI_{na-mb}$ process are characterized by a large group velocity mismatch $|D_v|$, therefore the phase-matching condition is typically dominated by the linear term $L$. We may therefore rewrite $\Omega_{PK}$, which solves $L(\Omega_{PK})+NL=0$, as the sum of a linear contribution $\Omega_{PK,L}$, which solves $L(\Omega_{PK,L})=0$, and a nonlinear contribution $\Omega_{PK,NL}<<\Omega_{PK,L}$, namely $\Omega_{PK}$=$\Omega_{PK,L}+\Omega_{PK,NL}$.\\
The solution for $\Omega_{PK,L}$ of $L=0$ depends on the sign of $D_{\beta 3}$,$\bar{\beta_{2}}$ and $D_v$. If $sign(D_{\beta 3})$=$sign(D_v)$=$-sign(\bar{\beta_{2}})$ then we find the 2 distinct positive values $\Omega_{PK,L}^{(1,2)}$. Otherwise, if $sign(D_{\beta 3})$=-$sign(D_v)$, then we find 1 positive solution $\Omega_{PK,L}^{(1)}$ and if $sign(D_{\beta 3})$=$sign(D_v)$=$sign(\bar{\beta_{2}})$ then no positive solutions are found.\\
We can easily show that the dynamics for the $MI_{ina-smb}$ process is reversed: if $sign(D_{\beta 3})$=$sign(D_v)$=$-sign(\bar{\beta_{2}})$ then no positive solutions are found, whereas if $sign(D_{\beta 3})$=-$sign(D_v)$ we find 1 positive solution $\Omega_{PK,L}^{(2)}$ and if $sign(D_{\beta 3})$=$sign(D_v)$=$sign(\bar{\beta_{2}})$ then we find the 2 distinct positive solutions $\Omega_{PK,L}^{(1,2)}$.\\
Therefore, overall 2 distinct solutions $\Omega_{PK,L}^{(1,2)}$ are found for the $MI_{na-mb}$ process, which can be associated to the $MI_{sna-imb}$ or the $MI_{ina-smb}$ processes. They are reported in Eq.(\ref{NaMbSpectrum}).\\
In order to calculate $\Omega_{PK,NL}$ we write $\Omega=\Omega_{PK,L}+\Omega_{PK,NL}$ and we expand the powers $\Omega^2$ and $\Omega^3$ of $L$ neglecting terms in $\Omega_{PK,NL}^2$ and $\Omega_{PK,NL}^3$. In this way $L+NL=0$ becomes a linear equation for $\Omega_{PK,NL}$, whose solution is reported in Eq.(\ref{NaMbSpectrum}). By means of $\Omega_{PK,NL}$ and $\Omega_{PK,L}$ we provide a fully analytical estimation of $\Omega_{PK}$.\\
Typically the bandwidth of $B_{na-mb}$ is small if compared to $\Omega_{PK}$, that is $|\Omega-\Omega_{PK}|<<\Omega_{PK}$, and we can thus expand ${\bf M'}[1,1]-{\bf M'}[2,2]$$\equiv$$L(\Omega)+NL$ about $\Omega_{PK}$. The derivative $\partial L/\partial\Omega$ evaluated at $\Omega_{PK}$ is $3D_{\beta_3}\Omega_{PK}^2 + 2\bar{\beta_2}\Omega_{PK} + D_v \approx -2k_2 \Omega_{PK,NL}^{-1}$ (see fourth line of Eq.(\ref{NaMbSpectrum}) setting $\Omega_{PK,L}\approx \Omega_{PK}$), with $k_2=(b_S/2) (C_{nn}|p_{na}|^2 + C_{mm}|p_{mb}|^2)$. The first order expansion of $L(\Omega)+NL$ is therefore  $L(\Omega_{PK})+NL -2 k_2 \Omega_{PK,NL}^{-1}(\Omega - \Omega_{PK})= -2 k_2 \Omega_{PK,NL}^{-1}(\Omega - \Omega_{PK})$, and in Eq.(\ref{app2_1}) we thus approximate $({\bf M'}[1,1]-{\bf M'}[2,2])^2\equiv (L+NL)^2 \approx 4 k_2^2 \Omega_{PK,NL}^{-2}(\Omega - \Omega_{PK})^2$, which results in the estimate of $B_{na-mb}$ stated in Eq.(\ref{NaMbSpectrum}).

\end{document}